\documentclass[11pt]{article}
\usepackage{a4wide}
\usepackage{color}

\usepackage[bookmarks=true,bookmarksnumbered=true,setpagesize=false]{hyperref}

\usepackage{enumerate}
\usepackage{amsmath,amssymb,amsthm}
\usepackage{textcomp,mathcomp} 

\usepackage{graphicx}
\usepackage{braket}
\usepackage{url}

\usepackage{cancel}

\usepackage{dsfont}
\newcommand{\Xds}{\mathds X}
\newcommand{\Pds}{\mathds P}

\usepackage{mathrsfs}

\newcommand{\wh}[1]{\widehat{#1}}

\newcommand{\tx}{\text}

\usepackage{comment}

\newcommand{\bp}{\begin{pmatrix}}
\newcommand{\ep}{\end{pmatrix}}
\newcommand{\bb}{\begin{bmatrix}}
\newcommand{\eb}{\end{bmatrix}}

\DeclareMathOperator{\Tr}{Tr}

\newcommand{\Ex}[2]{\mathbb E\!\left\{#1\right\}_{#2}}

\newcommand{\df}{\text{d}}

\newcommand{\al}[1]{\begin{align}#1\end{align}}
\newcommand{\als}[1]{\begin{align*}#1\end{align*}}
\newcommand{\ol}{\overline}

\newcommand{\ab}[1]{\left|#1\right|}
\newcommand{\abb}[1]{\left\|#1\right\|}
\newcommand{\paren}[1]{\left(#1\right)}
\newcommand{\pn}[1]{\left(#1\right)}
\newcommand{\sqbr}[1]{\left[#1\right]}
\newcommand{\br}[1]{\left\{#1\right\}} 

\newcommand{\autospace}{%
  \mathchoice%
    {\!}
    {\!}
    {}
    {}
}
\newcommand{\fn}[1]{\autospace\paren{#1}} 
\newcommand{\fnl}[1]{\autospace\sqbr{#1}} 
\newcommand{\Fn}[1]{\autospace\Pn{#1}} 
\newcommand{\Fnl}[1]{\autospace\Sqbr{#1}} 

\newcommand{\Ab}[1]{\bigl|#1\bigr|}
\newcommand{\Abb}[1]{\bigl\|#1\bigr\|}

\newcommand{\Pn}[1]{\bigl(#1\bigr)}
\newcommand{\Sqbr}[1]{\bigl[#1\bigr]}
\newcommand{\Br}[1]{\bigl\{#1\bigr\}} 

\newcommand{\wt}{\widetilde}
\newcommand{\nn}{\nonumber\\}

\newcommand{\p}{\partial}

\newcommand{\sr}{\stackrel}

\newcommand{\red}[1]{{\color[cmyk]{0,0.8,1,0}#1}}

\newcommand{\blue}[1]{{\color[cmyk]{1,0.5,0,0}#1}}
\newcommand{\cyan}[1]{{\color[cmyk]{0.8,0,0,0}#1}}
\newcommand{\magenta}[1]{{\color[cmyk]{0.1,0.7,0,0}#1}}

\newcommand{\black}[1]{{\color{black}#1}}

\usepackage{fancybox}

\usepackage{amsthm}
\theoremstyle{definition}

\newcommand{\ov}{\over}

\newcommand{\commutator}[2]{\left[#1,\,#2\right]}
\newcommand{\anticommutator}[2]{\left\{#1,\,#2\right\}}

\newcommand{\mc}{\mathcal}

\newcommand{\la}[1]{#1^\star} 

\usepackage[OMLmathsfit]{isomath}

\newcommand{\pd}[1]{\cyan{\mathsfit{#1}}} 
\newcommand{\pdf}[2]{\cyan{\mathsfit{#1}\fn{\black{\text{$#2$}}}}} 

\usepackage{bm} 

\usepackage{xparse}
\NewDocumentCommand{\opf}{m O{} m}{\red{{#1}^{#2}\fn{\black{\text{$#3$}}}}}
\NewDocumentCommand{\op}{m O{}}{\red{\wh{#1}^{#2}}}
\NewDocumentCommand{\opnh}{m O{} O{}}{\red{{#1}^{#2}_{#3}}} 
\NewDocumentCommand{\opff}{m O{} O{} m}{\red{#1^{#2}_{#3}\fn{\black{\text{$#4$}}}}}

\NewDocumentCommand{\rff}{m O{} m}{\blue{#1^{#2}\fn{\black{\text{$#3$}}}}}

\NewDocumentCommand{\stf}{m O{} m}{\magenta{\mathsfit{#1^{#2}\fn{\black{\text{$#3$}}}}}} 
\NewDocumentCommand{\st}{m O{}}{\magenta{\mathsfit{#1^{#2}}}} 

\newcommand{\opa}[1]{\wh{#1}} 

\newcommand{\rf}[1]{\blue{#1}} 

\newcommand{\id}{\red{\wh1}} 

\NewDocumentCommand{\idf}{O{}}{\blue{\text{id}^{#1}}}
\NewDocumentCommand{\idff}{O{} m}{\blue{\text{id}^{#1}\fn{\black{\text{$#2$}}}}}

\newcommand{\pc}[1]{\mathsf{#1}} 
\newcommand{\M}{\pc{M}}

\newcommand{\J}{\pc{J}}

\newcommand{\Xin}{X_\tx{in}}
\newcommand{\Pin}{P_\tx{in}}
\newcommand{\sigmain}{\sigma_\tx{in}}
\newcommand{\rhoin}{\rh_\tx{in}}
\newcommand{\rhoout}{\rh_\tx{out}}

\newcommand{\sigmasum}{\sigma_\tx{sum}}
\newcommand{\sigmared}{\sigma_\tx{red}}

\newcommand{\ub}{\underbrace}

\newcommand{\ot}{\otimes}

\newcommand{\rh}{{\wh\rho}}
\newcommand{\sh}{{\wh\sigma}}
\newcommand{\Pih}{{\wh\Pi}}
\newcommand{\xih}{{\wh\xi}}

\newcommand{\Eh}{\wh E}
\newcommand{\Oh}{\wh O}
\newcommand{\nh}{\wh n}
\newcommand{\Nh}{\wh N}
\newcommand{\Ah}{\wh A}
\newcommand{\Bh}{\wh B}

\newcommand{\R}{\mathbb R}

\begin{document}
\title{
Gaussian Formalism: Joint Measurement\\
for Heisenberg's Uncertainty Relation for Errors\\
by Squeezed Coherent States
}
\author{Kin-ya Oda\thanks{E-mail: \tt odakin@lab.twcu.ac.jp} \mbox{} and Naoya Ogawa\thanks{E-mail: \tt naoyaogawa.phys@gmail.com}\bigskip\\
\it\normalsize
$^*$Department of Information and Mathematical Sciences, \\
\it\normalsize Tokyo Woman's Christian University, Tokyo 167-8585, Japan\\
\it\normalsize
$^\dagger$Department of Complex Systems Science, Graduate School of Informatics,\\
\it\normalsize Nagoya University, Nagoya 464-8601, Japan
}
\maketitle
\begin{abstract}\noindent
We point out that the Gaussian wave-packet formalism can serve as a concrete realization of the joint measurement of position and momentum, which is an essential element in understanding Heisenberg’s original philosophy of the uncertainty principle, in line with the universal framework of error, disturbance, and their uncertainty relations developed by Lee and Tsutsui. We show that our joint measurement in the Gaussian phase space, being a Positive-Operator-Valued-Measure (POVM) measurement, smoothly interpolates between the projective measurements of position and momentum. We, for the first time, have obtained the Lee-Tsutsui (LT) error and the refined Lee error for the position-momentum measurement. We find that the LT uncertainty relation becomes trivial, $0=0$, in the limiting case of projective measurement of either position or momentum. Remarkably, in contrast to the LT relation, the refined Lee uncertainty relation, which assesses errors for local representability, provides a constant lower bound unaffected by these limits and is invariably saturated, for a pure Gaussian initial state. The obtained lower bound is in agreement with Heisenberg's value.
\end{abstract}

\newpage

\section{Introduction}
Heisenberg's Uncertainty Principle, introduced in 1927, is a cornerstone of quantum mechanics, fundamentally altering our understanding of measurement precision and predictability at the quantum level~\cite{Heisenberg:1927zz}. Originally, this principle concerned the errors and disturbances in the measurement of physical observables. This pivotal concept was then mathematically refined by Kennard, enhancing the principle's precision~\cite{Kennard:1927vgq}, and a modern alternative proof using the Cauchy-Schwarz inequality was provided by Weyl~\cite{Weyl1928}, before Robertson generalized the result to apply to any pair of observables~\cite{Robertson:1929zz}. The result of these developments is now known as the Kennard-Robertson (KR) inequality; see also Ref.~\cite{Schrodinger:1930ty} for Schr\"odinger's subsequent contribution, now known as the Schr\"odinger inequality. Unlike the original Heisenberg's principle, the KR and Schr\"odinger inequalities exclusively concern the fluctuations of physical observables, and are completely decoupled from the errors and disturbances arising from the measurements.

Building upon this foundational research, the contributions of Arthurs, Kelly, and Goodman led to the formulation of what is now known as the AKG inequality, significantly advancing our understanding of errors and the costs of measurements in quantum mechanics~\cite{Arthurs:1965eyx,PhysRevLett.60.2447}. Ozawa further expanded upon these concepts, resulting in the development of the Ozawa inequality through his studies on error-disturbance and error relations~\cite{PhysRevA.67.042105,OZAWA2004367}. Estimation theory, as highlighted in the works of Yuen-Lax~\cite{Yuen:1973mjw} and further developed by Watanabe-Sagawa-Ueda, culminated in the WSU inequality~\cite{PhysRevA.84.042121,watanabe2011quantum}, offering alternative perspectives. These collective advancements have significantly enhanced our comprehension of quantum uncertainty.


Lee and Tsutsui have provided an operationally tangible uncertainty relation, which we call the Lee-Tsutsui (LT) inequality, marking a significant advancement in the field of quantum uncertainty~\cite{lee2020geometric,lee2023universalformulationuncertaintyrelation,LEE2024129962}. The LT inequality offers a universal and geometric perspective on quantum measurement errors and disturbances, unifying the KR, Schr\"odinger, AKG, Ozawa, and WSU inequalities into a comprehensive, operationally interpretable framework. This framework was subsequently applied to a concrete two-state system~\cite{Lee:2020apk}.
Lee further expanded this approach by introducing the concepts of local representability and joint measurability, leading to the establishment of what we call the Lee inequality~\cite{Lee:2022cho,Lee:2022zgt}.
However, a challenge of this elegant framework is the lack of a concrete realization for the abstractly constructed mathematical objects, beyond the simple two-state system presented in Ref.~\cite{Lee:2020apk}.
The current paper aims to fill this gap, by providing a concrete realization of the key object in Lee's inequality, the representability for a joint measurement, in the case of Heisenberg's position-momentum uncertainty.

Recent developments in the Gaussian wave-packet formalism have opened up new possibilities in quantum field theory (QFT)~\cite{Ishikawa:2005zc,Ishikawa:2018koj,Ishikawa:2020hph,Mitani:2023hpd,Ishikawa:2023bnx}. Particularly notable is the rigorous proof of the emergence of the time-boundary effect, initially claimed in Ref.~\cite{Ishikawa:2013kba}, which has been recently established~\cite{Ishikawa:2021bzf}.
Gaussian wave packets are significant as they form an (over)complete set that spans the position-momentum phase space, including that of the free one-particle subspace, which underpins ordinary non-relativistic quantum mechanics~\cite{Ishikawa:2005zc}; see also Refs.~\cite{PhysRev.130.2529,Kaiser:1977ys,Kaiser:1978jg,Oda:2021tiv} for a viewpoint that the Gaussian basis can be regarded as a complete set of squeezed coherent states in position-momentum space.\footnote{
We refer to these position-momentum coherent states as the ``Gaussian wave packets,'' and the complete set formed by them as the ``Gaussian basis,'' conforming to the QFT literature, even though the main focus of this paper is quantum mechanics in the one-particle subspace on a fixed time slice. This is because they can be naturally extended to the spacetime in QFT, as discussed in the main text. (It is worth noting that, while the primary focus of Glauber's pioneering work~\cite{PhysRev.130.2529} is on coherent states in field space for quantum optics, it already addresses the coherent states in position-momentum space.)
}
Furthermore, the Gaussian basis can be generalized to a complete set of Lorentz-invariant wave-packet basis~\cite{Kaiser:1977ys,Kaiser:1978jg,Oda:2021tiv}, and further to Lorentz-covariant wave-packet basis for spinor fields, incorporating spin degrees of freedom~\cite{Oda:2023qek}.

The importance of the phase space is also emphasized in Refs.~\cite{2013PhRvL.111p0405B,Busch:2013vba,Busch:2014mkj}, particularly in the context of providing a rigorous examination and proof of Heisenberg-type inequalities, with a focus on the precision-disturbance trade-off in quantum measurements.

In this paper, we posit that a Positive-Operator-Valued-Measure (POVM) measurement onto the Gaussian wave-packet basis naturally serves as the joint measurement for position and momentum, a key element in the discussion on Heisenberg's uncertainty in Ref.~\cite{Lee:2022zgt}. We provide concrete expressions for the pullback and pushforward, which remained abstract in the original construction~\cite{lee2020geometric,LEE2024129962,Lee:2020apk,Lee:2022cho,Lee:2022zgt}. We will demonstrate that the Gaussian basis naturally interpolates between the position-space basis and the momentum-space basis in the limits of infinitely narrow and wide widths, respectively.

The organization of this paper is as follows:
In Sec.~\ref{Lee-Tsutsui section}, we briefly review the Lee-Tsutsui (LT) formalism to clarify our notation, focusing on the parts relevant to our discussion.
In Sec.~\ref{Gaussian section}, we review the Gaussian formalism and show that the Gaussian wave packet smoothly interpolates between the position and momentum bases.
In Sec.~\ref{separate measurement section}, we study the separate measurement of position or momentum of a Gaussian initial state in the LT formalism. We show the concrete forms of the pullback and pushforward for the position and momentum, which were previously only abstract constructs.
In Sec.~\ref{Gaussian joint measurement section}, we present our main findings: that the POVM measurement onto the Gaussian basis can be regarded as the joint measurement of position and momentum. Again we show the concrete form of the pullback and pushforward for the position and momentum for the joint measurement of the Gaussian initial state.
In Sec.~\ref{summary section}, we summarize our results and discuss future directions.
In Appendix~\ref{Ozawa section} and \ref{Ozawa and LT section}, we briefly review the Ozawa inequality and its relationship to the LT inequality, respectively.

\section{Lee-Tsutsui formalism}\label{Lee-Tsutsui section}
We briefly review the formalism of Lee and Tsutsui~\cite{lee2020geometric,LEE2024129962,Lee:2020apk} and of Lee~\cite{Lee:2022cho,Lee:2022zgt}, which we collectively call the LT formalism hereafter, focusing on the part relevant to our discussion.

\begin{table}\centering
%
%
%
\begin{tabular}{c|clcl}
&\multicolumn{2}{l}{Observables} & \multicolumn{2}{l}{States}\\
\hline
Quantum&$\opff S{\mc H}$&$\ni\op A$ (self-adjoint operator)&$\stf Z{\mc H}$&$\ni\st\rh$ (density operator)\\
Classical&$\rff R\Omega$&$\ni\rf f$ (real function)&$\pdf W\Omega$&$\ni\pd p$ (PDF)\\
\end{tabular}
\caption{Reminder table for the spaces. A typical element is also given for each space.}\label{spaces table}
\end{table}

\subsection{Basic quantum mechanics}
A physical \emph{observable} is represented by a \emph{self-adjoint operator} $\op A$ that satisfies $\op A[\dagger]=\op A$, where $\op A[\dagger]$ is the adjoint of $\op A$, on a Hilbert space $\mc H$. We write the linear space of all the self-adjoint operators as $\opff S{\mc H}$; see Table~\ref{spaces table}.
A physical \emph{state} is represented by a \emph{density operator} $\st\rh$ that is self-adjoint, $\st\rho[\dagger]=\st\rho$, has a trace of unity, $\Tr\fnl{\st\rh}=1$, and has a spectrum that is entirely non-negative, $\st\rh\geq0$. We write the state space $\stf Z{\mc H}$; see Table~\ref{spaces table}. Typically in our discussion, one may assume the state to be a \emph{pure state} given as a \emph{projection} operator $\st\rh=\Ket{\psi}\Bra{\psi}$, where $\Ket{\psi}\in\mc H$ (and $\Bra{\psi}$ is its dual). With this case in mind, we also call a vector $\Ket{\psi}\in\mc H$ the state.

An expectation value of (not necessarily self-adjoint) operator $\op A$ on $\st\rh$ is given by
\al{
\Braket{\op A}_{\st\rh}
	&:=	\Tr\fnl{\op A\st\rh},
		\label{expectation value}
}
and a semi-inner product of any (not necessarily self-adjoint) operators $\op A$ and $\op B$ by
\al{
\Braket{\op A,\,\op B}_{\st\rh}
	&:=	\Braket{\anticommutator{\op A}{\op B}\ov2}_{\st\rh},
}
where the commutator and anticommutator are $[\op A,\op B]:=\op A\op B-\op B\op A$ and $\{\op A,\op B\}:=\op A\op B+\op B\op A$, respectively. The semi-inner product naturally induces the semi-norm and the standard deviation for any (not necessarily self-adjoint) operator $\op A$ on $\st\rh$:\footnote{
In this paper, we regard a map from a space to $\mathbb R$ (or $\mathbb C$) as a \emph{functional}, and place its argument into square brackets rather than into parentheses, conforming to physics community notation.
}
\al{
\abb{\op A}_{\st\rh}
	&:=	\sqrt{\Braket{\op A[\dagger],\,\op A}_{\st\rh}},&
\sigma_{\st\rh}\fnl{\op A}
	&:=	\sqrt{\abb{\op A}_{\st\rh}^2-\Braket{\op A}_{\st\rh}^2}.
		\label{quantum standard deviation}
}
If $\op A$ is self-adjoint, the standard deviation reduces to $\sigma_{\st\rh}^2\Fnl{\op A}=\braket{\op A[2]}_{\st\rh}-\braket{\op A}_{\st\rh}^2$.
Further if $\st\rh$ is a pure state $\st\rh=\Ket\psi\Bra\psi$, the semi-norm reduces to the norm of the state after the operation, $\|\op A\|_{\Ket{\psi}\Bra{\psi}}^2=\|\op A\Ket\psi\|^2$, and accordingly the expectation value~\eqref{expectation value} reduces to $\braket{\op A}_{\Ket{\psi}\Bra{\psi}}=\Bra{\psi}\op A\Ket{\psi}$.

\subsection{Measurement process}
A \emph{measurement} $\M$ is an \emph{affine map} $\st\rh\mapsto\pd p$ in the sense that, for any states $\st{\rh_1},\st{\rh_2}$ and $\forall\lambda\in[0,1]$,
\al{
\M\Pn{\lambda\st{\rh_1}+\pn{1-\lambda}\st{\rh_2}}
	&=	\lambda\M\st{\rh_1}+\pn{1-\lambda}\M\st{\rh_2},
		\label{affinity}
}
where $\pd p$ is a \emph{probability density function} (PDF) that satisfies $\int_\Omega\df\omega\,\pdf p{\omega}=1$ and, $\forall\omega\in\Omega$, $\pdf p{\omega}\geq0$, in which $\Omega$ is a \emph{sample (or outcome) space} and $\df\omega$ its measure. We write the convex set of all the PDFs as $\pdf W{\Omega}$; see Table~\ref{spaces table}. Hereafter, we write $\pd{M\rh}:=\M\st\rh$ as a reminder to reinforce that $\pd{M\rh}$ is a PDF. As a whole, $\M\colon\stf Z{\mc H}\to\pdf W\Omega$ is defined.

Two comments are in order:
First, the affinity~\eqref{affinity} is the least requirement to allow the interpretation of probability mixture but is powerful enough to derive most of the results in the LT formalism.
An affine map between any pair of state spaces, in the right of Table~\ref{spaces table}, is in general called a \emph{process}.
The LT formalism can handle all the processes on equal foot.
Especially when applied to a \emph{quantum process} $\pc\Theta\colon\stf{Z_1}{\mc H_1}\to\stf{Z_2}{\mc H_2}$, its \emph{disturbance} can be described in quite a parallel manner to various $\M$'s \emph{errors} given below.
For notational simplicity, we do not exploit this beauty of the formalism, and will stick to a measurement $\M$ as a process (except in Sec.~\ref{joint measurement section}).
Second, for most of our purpose, one may assume a \emph{projective measurement} that gives, $\forall\st\rh\in\stf Z{\mc H}$,\footnote{
Here, the term ``projective'' should be understood in a broader sense; see footnote~\ref{projective footnote}.
}
\al{
\pdf{\sqbr{M\rh}}{\omega}=\Tr\Fnl{\Ket{\omega}\Bra{\omega}\st\rh}.
	\label{projective measurement}
}

On the sample space $\Omega$, we define the \emph{classical} expectation value, semi-inner product, semi-norm, and standard deviation in accordance with their quantum counterparts:
\al{
\Braket{\rf f}_{\pd p}
	&:=	\int_\Omega\df\omega\,\rff f{\omega}\pdf p{\omega},&
		\label{classical expectation value}
\Braket{\rf f,\,\rf g}_{\pd p}
	&:=	\int_\Omega\df\omega\,\rff f[*]{\omega}\rff g{\omega}\pdf p{\omega},\\
\abb{\rf f}_{\pd p}
	&:=	\sqrt{\Braket{\rf f,\,\rf f}_{\pd p}}
	=	\sqrt{\int_\Omega\df\omega\ab{\rff f{\omega}}^2\pdf p{\omega}},&
\sigma_{\pd p}\fnl{\rf f}
	&:=	\sqrt{\abb{\rf f}_{\pd p}^2-\ab{\Braket{\rf f}_{\pd p}}^2},
		\label{classical standard deviation}
}
where $\rf f,\rf g$ are arbitrary real functions
on $\Omega$ (but we have retained the complex conjugate, denoted by $*$ per physics community notation, to accommodate possible generalizations). More concretely for the real function $\rf f$, we have
$\sigma_{\pd p}^2\fnl{\rf f}=\int_\Omega\df\omega\rff f[2]\omega\pdf p\omega-\pn{\int_\Omega\df\omega\rff f\omega\pdf p\omega}^2$.

\subsection{LT adjoint, pullback, and pushforward}
Given a measurement $\M$, we define (what we call) the \emph{Lee-Tsutsui (LT) adjoint} operator $\op{\la Mf}$ of a real function $\rf f$ by, $\forall\st\rh\in\stf Z{\mc H}$,
\al{
\Braket{\op{\la Mf}}_{\st\rh}
	&=	\Braket{\rf f}_{\pd{M\rh}}.
		\label{LT adjoint defined}
}
We also call this the \emph{pullback} operator of $\rf f$.\footnote{\label{footnote for semi-ness}
In the original construction~\cite{lee2020geometric,LEE2024129962,Lee:2020apk,Lee:2022cho,Lee:2022zgt}, it was crucial to define the pullback (and pushforward given below) \emph{locally} in the sense that, everything is partitioned into equivalence classes as $[\op{\la Mf}]_{\st\rh}$, $\sqbr{\rf f}_{\pd p}$, etc., by the equivalences
\als{
\op A\sr{\st\rh}\sim\op B
	&\Longleftrightarrow\abb{\op A-\op B}_{\st\rh}=0,&
\rf f\sr{\pd p}\sim\rf g
	&\Longleftrightarrow\abb{\rf f-\rf g}_{\pd p}=0.
}
For our purpose, this step is not necessary and we neglect it, though can be important for other purposes.
}
As a whole, we define the LT adjoint $\la\M\colon\rff R\Omega\to\opff S{\mc H}$, where $\rff R\Omega$ is the space of all the real functions on $\Omega$; see Table~\ref{spaces table}.

It is important that the following inequality follows from the Kadison-Schwarz inequality:
\al{
\abb{\rf f}_{\pd{M\rh}}\geq\abb{\op{\la Mf}}_{\st\rh}.
	\label{inequality for LT adjoint}
}
By definitions~\eqref{quantum standard deviation}, \eqref{classical standard deviation}, and \eqref{LT adjoint defined}, we may rewrite inequality~\eqref{inequality for LT adjoint} into that of the standard deviations:
\al{
\sigma_{\pd{M\rh}}\fnl{\rf f}
	&\geq
		\sigma_{\st\rh}\fnl{\op{\la Mf}}.
}
To quote~\cite{Lee:2022zgt}, \emph{the operational cost of acquiring the expectation value of a quantum observable through measurements can never break the quantum limit imposed by the said observable.}

Once the LT adjoint is obtained, we define the \emph{pushforward} function $\rf{M_\star\opa A}$ of a given self-adjoint operator~$\op A$ (over a state $\st\rh$) by, $\forall\rf f\in\rff R\Omega$,
\al{
\Braket{\op A,\,\op{\la M f}}_{\st\rh}
	&=	\Braket{\rf{M_\star\Ah},\,\rf f}_{\pd{M\rh}},
	\label{definition of pushforward}
}
or more concretely by, $\forall\rf f\in\rff R\Omega$,
\al{
\Tr\fnl{{\anticommutator{\op A}{\op{\la M f}}\ov2}\st\rh}
&=
\int_\Omega\df\omega\,
	\rff{\sqbr{M_\star\Ah}}\omega\,
	\rff f\omega\,
	\pdf{\sqbr{M\rh}}\omega.
	\label{concrete definition of pushforward}
}
By construction, the pushforward function is obtained \emph{locally}, namely obtained for each given~$\st\rh$, and should be more properly written as $\rf{M_{\rh\star}\opa A}$. However, this dependence on $\st\rh$ is trivially understood for most of our purposes, and we will use the shorthand notation to avoid clutter. As a whole, we define the pushforward $\M_\star\colon\opff S{\mc H}\to\rff R\Omega$ (for each given $\st\rh$ as mentioned).

We note that for the case of projective measurement~\eqref{projective measurement}, its LT adjoint also becomes projective, in the sense that given any function $\rf f\in\rff R\Omega$,
\al{
\op{\la Mf}=\int_\Omega\df\omega\rff f\omega\Ket{\omega}\Bra{\omega}.
	\label{projective LT adjoint}
}
Also, the pushforward for the projective measurement of $\op A$ becomes the identity $\rf{M_\star\opa A}=\idf$; see Eq.~\eqref{concerete realization of id} below for a concrete realization.

For an operator $\op A$ on $\mc H$ and a state $\st\rh\in\stf Z{\mc H}$, the following inequality can be shown in parallel to Eq.~\eqref{inequality for LT adjoint}:
\al{
\abb{\rf{M_\star\wh A}}_{\pd{M\rh}}
	\leq	\abb{\op A}_{\st\rh}
	\leq	\abb{\rf{M^{\star-1}\wh A}}_{\pd{M\rh}},
		\label{inequalities for pushforward and inverse}
}
where
$\rf{M^{\star-1}\opa A}
	:=	\pn{\la\M}^{-1}\op A$.\footnote{
When the inverse cannot be defined, a \emph{partial inverse} is introduced in Ref.~\cite{Lee:2022zgt}. For our purposes, this is an unnecessary detour, and we will neglect it, though it may be important for other contexts.
}
In what follows, the left and right inequalities will be used to define the LT and Lee errors, respectively.

\subsection{LT error and inequality}
We define (what we call) the LT error~\cite{lee2020geometric,LEE2024129962}:
\al{
\varepsilon_{\st\rh}\fnl{\op A;\M}
	&:=	\sqrt{\abb{\op A}^2_{\st\rh}
		-\abb{\rf{M_\star\Ah}}^2_{\pd{M\rh}}}
	=	\sqrt{\Tr\fnl{\op A[2]\st\rh}
		-\int_\Omega\df\omega\,
			\rff{\sqbr{M_\star\Ah}}[2]{\omega}\,
			\pdf{\sqbr{M\rh}}{\omega}
			},
		\label{LT error defined}
}
where $\rff{\Sqbr{M_\star\Ah}}[2]{\omega}:=\Pn{\rff{\Sqbr{M_\star\Ah}}{\omega}}^2$ as above.
The meaning of LT error is well explained in Ref.~\cite{lee2020geometric,LEE2024129962}, which we will briefly review hereafter.
Given an estimator function $\rf f\in\rff R\Omega$, we introduce an error with respect to $\rf f$ (\emph{abbr.} $\rf f$-error):
\al{
\varepsilon_{\st\rh}\fnl{\op A;\M,\rf f}
	&:=	\sqrt{\abb{\op A-\op{\la M f}}^2_{\st\rh}
		+\pn{
			\abb{\rf f}_{\pd{M\rh}}^2
			-\abb{\op{\la M f}}_{\st\rh}^2
			}}.
			\label{f-error}
}
The first term in the square root is the goodness of the fit of $\op A$ by the pullback operator of the estimator $\rf f$, while the second is the increase in the variance from that of the quantum pullback operator~$\op{\la M f}$ to that of the classical estimator $\rf f$:
\al{
\abb{\rf f}_{\pd{M\rh}}^2
		-\abb{\op{\la M  f}}_{\st\rh}^2
	&=	\pn{
			\abb{\rf f}_{\pd{M\rh}}^2
			-\Braket{\rf f}_{\pd{M\rh}}^2
			}
		-\pn{
			\abb{\op{\la M  f}}_{\st\rh}^2
			-\Braket{\op{\la M  f}}_{\st\rh}^2
			}\nn
	&=	\sigma_{\pd{M\rh}}^2\fnl{\rf f}
		-\sigma_{\st\rh}^2\fnl{\op{\la M f}},
		\label{2nd term}
}
where we used the definition of LT adjoint~\eqref{LT adjoint defined}.

The first term in the square root in Eq.~\eqref{f-error} reads
\al{
\abb{\op A-\op{\la M f}}^2_{\st\rh}
	&=	\Braket{\op A,\,\op A}_{\st\rh}
		-2\Braket{\op A,\,\op{\la M f}}_{\st\rh}
		+\Braket{\op{\la M f},\,\op{\la M f}}_{\st\rh}\nn
	&=	\abb{\op A}_{\st\rh}^2
		-2\Braket{\rf{M_\star\Ah},\,\rf f}_\pd{M\rh}
		+\abb{\op{\la M f}}_{\st\rh}^2,
		\label{1st term}
}
where we used the definition of pushforward~\eqref{definition of pushforward}.
Putting Eq.~\eqref{1st term} into Eq.~\eqref{f-error}, we obtain
\al{
\varepsilon_{\st\rh}^2\fnl{\op A;\M,\rf f}
	&=	\varepsilon_{\st\rh}^2\fnl{\op A;\M}
		+\abb{\rf{M_\star \Ah}-\rf f}_{\pd{M\rh}}^2.
		\label{f-error and LT error}
}
The $\rf f$-error is minimized by choosing $\rf f$ to be the optimal function $\rf{M_\star\Ah}$, and the LT error~\eqref{LT error defined} gives the smallest error achieved by this choice.

The LT error satisfies (what we call) the LT inequality:
\al{
\varepsilon_{\st\rh}\fnl{\op A;\M}
\varepsilon_{\st\rh}\fnl{\op B;\M}
	\geq \sqrt{
		\mc I_{\st\rh}^2\fnl{\op A,\op B;\M}
		+\mc R_{\st\rh}^2\fnl{\op A,\op B;\M}
		},
		\label{LT inequality}
}
where
\al{
\mc I_{\st\rh}\fnl{\op A,\op B;\M}
	&:=	\Braket{\commutator{\op A}{\op B}\ov2i}_{\st\rh}
		-\Braket{
			\commutator{\op{\la M  M_\star \Ah}}{\op B}\ov2i
			}_{\st\rh}
		-\Braket{\commutator{\op A}{\op{\la M  M_\star \Bh}}\ov2i}_{\st\rh},
			\label{imaginary part for LT bound}\\
\mc R_{\st\rh}\fnl{\op A,\op B;\M}
	&:=	\Braket{\anticommutator{\op A}{\op B}\ov2}_{\st\rh}
		-\Braket{
			\rf{M_\star \Ah},\,
			\rf{M_\star \Bh}
			}_{\pd{M\rh}}.
			\label{real part for LT bound}
}
The imaginary parts $\mc I_{\st\rh}\Fnl{\op A,\op B;\M}$ represents the quantum contribution to the lower bound, while the real parts $\mc R_{\st\rh}\Fnl{\op A,\op B;\M}$ the semi-classical contribution.

Given a semi-inner product and the corresponding norm,
\al{
\Braket{\pn{\op A,\rf f},\pn{\op B,\rf g}}_{\M,\st\rh}
	&:=	\Braket{\op A,\op B}_{\st\rh}
		+\Braket{\rf f,\rf g}_{\pd{M\rh}}
		-\Braket{\op{\la Mf},\op{\la Mg}}_{\st\rh},\\
\abb{\pn{\op A,\rf f}}_{\M,\st\rh}
	&:=	\sqrt{\Braket{\pn{\op A,\rf f},\pn{\op A,\rf f}}_{\M,\st\rh}},
}
the LT inequality~\eqref{LT inequality} is nothing but the Cauchy-Schwarz inequality
\al{
\abb{\pn{\op X_{\op A},\rf{f_{\Ah}}}}_{\M,\st\rh}\,
\abb{\pn{\op X_{\op B},\rf{f_{\Bh}}}}_{\M,\st\rh}
	&\geq
		\ab{\Braket{\pn{\op X_{\op A},\rf{f_{\Ah}}},\pn{\op X_{\op B},\rf{f_{\Bh}}}}_{\M,\st\rh}},
}
where
$\op X_{\op A}:=\op A-\op{\la MM_\star\Ah}$ and
$\rf{f_{\Ah}}:=\rf{M_\star\Ah}$.

\subsubsection{Relation between LT and Ozawa inequalities}
We comment on a relation of the LT inequality~\eqref{LT inequality} to the Ozawa inequality~\cite{OZAWA2004367}:
\al{
\epsilon^\tx O_{\st\rh}\fnl{\op A}\epsilon^\tx O_{\st\rh}\fnl{\op B}
+\epsilon^\tx O_{\st\rh}\fnl{\op A}\sigma_{\st\rh}\fnl{\op B}
+\sigma_{\st\rh}\fnl{\op A}\epsilon^\tx O_{\st\rh}\fnl{\op B}
	&\geq
		{1\ov2}\ab{\Braket{\commutator{\op A}{\op B}}_{\st\rh}},
		\label{Ozawa inequality}
}
where $\epsilon^\tx{O}_{\st\rh}$ is the Ozawa error, which is always larger than the LT error; see Appendix~\ref{Ozawa section} for a review.
The LT inequality~\eqref{LT inequality} is a necessary condition for the Ozawa inequality in the following sense:
\al{
\epsilon^\tx O_{\st\rh}\fnl{\op A}\epsilon^\tx O_{\st\rh}\fnl{\op B}
	\geq
		\varepsilon_{\st\rh}\fnl{\op A}
		\varepsilon_{\st\rh}\fnl{\op B}
	&\sr{\tx{LT}}\geq
		\sqrt{
		\mc I_{\st\rh}^2\fnl{\op A,\op B}
		+\mc R_{\st\rh}^2\fnl{\op A,\op B}
		}\nn
	&\geq
		\ab{\mc I_{\st\rh}\fnl{\op A,\op B}}
	\geq
		{1\ov2}\ab{\Braket{\commutator{\op A}{\op B}}_{\st\rh}}
		-\epsilon^\tx O_{\st\rh}\fnl{\op A}\sigma_{\st\rh}\fnl{\op B}
		-\sigma_{\st\rh}\fnl{\op A}\epsilon^\tx O_{\st\rh}\fnl{\op B},
\label{Ozawa and LT inequalities}
}
where we have abbreviated $\M$ from the errors and related quantities;
see Appendix~\ref{Ozawa and LT section} for its derivation.

\subsubsection{Impossibility of vanishing LT errors for non-commuting observables}
\label{Impossibility of vanishing LT errors for non-commuting observables}
We see from the Ozawa inequality~\eqref{Ozawa inequality} that, when $\op A$ and $\op B$ do not commute (i.e.\ when the right-hand side of Eq.~\eqref{Ozawa inequality} is non-zero), it is impossible for both Ozawa errors $\epsilon^\tx O_{\st\rh}\Fnl{\op A}$ and $\epsilon^\tx O_{\st\rh}\Fnl{\op B}$ to vanish, unless one considers a limit in which the standard deviation becomes infinite.
Let us now consider the LT inequality\,\eqref{LT inequality}.
As shown in Eq.~\eqref{LT inequality for Gaussian initial state}, the right-hand side of the LT inequality can indeed vanish.
However, this does not imply that both LT errors can vanish simultaneously for non-commuting observables.

To see why, we first note that the condition $\varepsilon_{\st\rh}\Fnl{\op A}=0$ is equivalent to~\cite{lee2020geometric,LEE2024129962}:
\al{
\op A=\op{\la M  M_\star \Ah}.
	\label{equality to pullback of pushforward}
}
This can be shown by taking $\rf f=\rf{M_\star \Ah}$ in Eq.~\eqref{f-error and LT error}, yielding
\al{
\varepsilon_{\st\rh}\fnl{\op A}
	=	\varepsilon_{\st\rh}\fnl{\op A;\M,\rf{M_\star \Ah}}
	=	\sqrt{\abb{\op A-\op{\la M  M_\star \Ah}}^2_{\st\rh}
		+\pn{
			\abb{\rf{M_\star \Ah}}_{\pd{M\rh}}^2
			-\abb{\op{\la M  M_\star \Ah}}_{\st\rh}^2
			}}.
}
As seen in Eq.~\eqref{inequalities for pushforward and inverse}, the subtraction in the above parentheses is always non-negative.
Hence, for $\varepsilon_{\st\rh}\Fnl{\op A}$ to vanish, we must have $\Abb{\op A-\op{\la M  M_\star \Ah}}_{\st\rh}=0$, implying $\op A=\op{\la M  M_\star \Ah}$.\footnote{
In a general setup, a vanishing semi-norm need not imply the operator itself vanishes. 
This subtlety is resolved by considering operators on the quotient spaces discussed in footnote~\ref{footnote for semi-ness}.
}
To show the converse, it suffices to note that the above subtraction vanishes if Eq.~\eqref{equality to pullback of pushforward} holds, by virtue of inequality~\eqref{inequalities for pushforward and inverse}.

Next, if $\varepsilon_{\st\rh}\Fnl{\op A}=\varepsilon_{\st\rh}\Fnl{\op B}=0$, then Eq.~\eqref{equality to pullback of pushforward} gives
\al{
\mc I_{\st\rh}\fnl{\op A,\op B;\M}
	&=	-\Braket{\commutator{\op A}{\op B}\ov2i}_{\st\rh},
}
whose square is non-zero for non-commuting observables.
Hence, the LT inequality\,\eqref{LT inequality} would reduce to a zero left-hand side being greater than a positive right-hand side.
Thus, the LT inequality forbids the simultaneous vanishing of both LT errors for non-commuting operators.

\subsection{Lee error and inequality}
Following Ref.~\cite{Lee:2022cho,Lee:2022zgt}, we define the \emph{error for local representability}, which we call the \emph{Lee error} hereafter:
\al{
\wt\varepsilon_{\st\rh}\fnl{\op A;\M}
	&:=	\sqrt{\abb{\rf{M^{\star-1}\opa A}}^2_\pd{M\rh}-\abb{\op A}^2_{\st\rh}}.
		\label{Lee error}
}
More explicitly,
\al{
\wt\varepsilon_{\st\rh}^2\fnl{\op A;\M}
	&=	\int_\Omega\df\omega\,\rff{\sqbr{M^{\star-1}\opa A}}[2]{\omega}\,\pdf{\sqbr{M\rh}}{\omega}
		-\Tr\fnl{\op A[2]\st\rh}.
}
It is worth noting that the $\rf f$-error~\eqref{f-error} reduces to the Lee error when the test function $\rf f=\rf{M^{\star-1}\opa A}$ is chosen:
\al{
\varepsilon_{\st\rh}\fnl{\op A;\M,\rf{M^{\star-1}\opa A}}
	&=	\wt\varepsilon_{\st\rh}\fnl{\op A;\M}.
}

The Lee inequality reads
\al{
\wt\varepsilon_{\st\rh}\fnl{\op A;\M}
\wt\varepsilon_{\st\rh}\fnl{\op B;\M}
	&\geq
		\sqrt{\mc I_{0\st\rh}^2\fnl{\op A,\op B}+\wt{\mc R}_{\st\rh}^2\fnl{\op A,\op B;\M}},
		\label{Lee inequality}
}
where the contributors to the lower bound are
\al{
\mc I_{0\st\rh}\fnl{\op A,\op B}
	&:=	\Braket{\commutator{\op A}{\op B}\ov2i}_{\st\rh},
		\label{imaginary for Lee inequality}\\
\wt{\mc R}_{\st\rh}\fnl{\op A,\op B;\M}
	&:=	\Braket{\anticommutator{\op A}{\op B}\ov2}_{\st\rh}-\Braket{\rf{M^{\star-1}\opa A},\,\rf{M^{\star-1}\opa B}}_{\pd{M\rh}}.
		\label{real for Lee inequality}
}
Again the imaginary and real parts represent quantum and semi-classical contributions, respectively.

\subsection{Joint measurement and classical projection of marginalization}\label{joint measurement section}
Following Lee's construction~\cite{Lee:2022cho,Lee:2022zgt}, we consider a joint measurement $\J\colon\stf Z{\mc H}\to\pdf W\Omega$, whose sample space (also called the outcome space) is separated into $\Omega=\Omega_1\times\Omega_2$ so that the outcome PDF is a function of $\omega_i\in\Omega_i$ ($i=1,2$), namely, $\pd p\in\pdf W\Omega$ is written as $\pdf p{\omega_1,\omega_2}$.
Now we introduce the following \emph{classical projection processes}~$\bm\pi_i\colon\pdf W{\Omega_1\times\Omega_2}\to\pdf W{\Omega_i}$ ($i=1,2$), defined by the projection to marginal:
\al{
\pdf{\sqbr{\pi_1p}}{\omega_1}
	&:=	\int_{\Omega_2}\df\omega_2\,\pdf p{\omega_1,\omega_2},&
\pdf{\sqbr{\pi_2p}}{\omega_2}
	&:=	\int_{\Omega_1}\df\omega_1\,\pdf p{\omega_1,\omega_2}.
		\label{marginalization}
}
If two measurements $\M_i\colon\stf Z{\mc H}\to\pdf W{\Omega_i}$ ($i=1,2$) can be written as $\M_i=\bm\pi_i\circ\J$, they are said to admit the joint measurement $\J$.

The LT adjoint $\la{\bm\pi}_i$ is given in parallel to that of the measurement~\eqref{LT adjoint defined}: Given a function $\rf f\in\rff R{\Omega_i}$, its LT adjoint function $\rf{\la \pi_if}\in\rff R{\Omega_1\times\Omega_2}$ is defined by, $\forall\pd p\in\pdf W{\Omega_1\times\Omega_2}$,
\al{
\Braket{\rf{\la \pi_if}}_{\pd p}
	&=	\Braket{\rf f}_{\pd{\pi_ip}},
		\qquad(i=1,2)
}
or more concretely, $\forall\pd p\in\pdf W{\Omega_1\times\Omega_2}$,
\al{
\int_{\Omega_1}\df\omega_1\int_{\Omega_2}\df\omega_2\,\rff{\sqbr{\la \pi_if}}{\omega_1,\omega_2}\,\pdf p{\omega_1,\omega_2}
	&=	\int_{\Omega_i}\df\omega_i\,\rff f{\omega_i}\,\pdf{\sqbr{\pi_ip}}{\omega_i}.
		\qquad(i=1,2)
}
We see that, by the definition~\eqref{marginalization}, the LT adjoint of $\rf f$ is trivially
\al{
\rff{\sqbr{\la \pi_if}}{\omega_1,\omega_2}
	&=	\rff f{\omega_i}.	\qquad(i=1,2)
		\label{adjoint of projection}
}

\section{Gaussian formalism}\label{Gaussian section}
We review the Gaussian wave-packet formalism within a free one-particle subspace, decoupled from the spin degrees of freedom. In this paper, we restrict our attention to the 1D position-momentum space since the generalization to higher dimensions is straightforward. Throughout the paper, we restrict ourselves to the consideration of a fixed time slice and do not consider the time evolution/translation; see  Refs.~\cite{Ishikawa:2005zc,Ishikawa:2018koj,Ishikawa:2020hph,Mitani:2023hpd} for a Gaussian formalism applied to a relativistic particle, Refs.~\cite{Kaiser:1977ys,Kaiser:1978jg,Oda:2021tiv} for insights into the free one-particle subspace of the relativistic particle and generalization to a Lorentz-invariant complete basis, and Ref.~\cite{Oda:2023qek} for an account on the transformation of the space-like hyperplane under the Lorentz transformation and for a generalization to include the relativistic spin degrees of freedom into a Lorentz-covariant complete basis.

\subsection{Plane-wave basis}
Let $\Set{\Ket{ p}}_{p\in\R}$ be the momentum-space basis that spans the 1D free one-particle subspace, which we write $\mc H$ hereafter, without spin degrees of freedom.
We employ the normalization
\al{
\Braket{p|p'}
	&=	\delta\fn{p-p'},
}
where the right-hand side is the Dirac delta function (distribution), resulting in the standard normalization for the completeness relation (resolution of identity):
\al{
\int_\R\df p\Ket{ p}\Bra{ p}
	&=	\id,
}
in which $\id$ is the identity operator on $\mc H$.
Here and hereafter, the integral region of momentum (and position) is always from $-\infty$ to $\infty$ unless otherwise stated.\footnote{\label{integral notation}
Usually one would write $\int_{-\infty}^\infty\df x$\, or $\int\df x$ in short. Here, we conform to the language of Eq.~\eqref{classical expectation value}.
}
The dual position-space basis $\Set{\Ket{ x}}_{x\in\R}$ is defined by
\al{
\Braket{ x| p}
	&:=	{e^{i p x}\ov\sqrt{2\pi}}.
		\label{x state defined}
}
Here and hereafter, we employ the natural unit $\hbar=1$, unless otherwise stated, so that momentum and wavenumber have the same dimension (unit).
The definition~\eqref{x state defined} leads to the completeness in the position space:
\al{
\int_\R\df x\Ket x\Bra x
	&=	\id.
}

We identify this position space as an arbitrarily chosen equal-time hyperplane.
It is worth reiterating that the position-space basis can be used to span any fixed time slice, even for a relativistic particle, both in quantum mechanics (see e.g.\ Ref.~\cite{Oda:2021tiv}) and in quantum field theory~\cite{Ishikawa:2005zc,Ishikawa:2018koj,Ishikawa:2020hph} (this viewpoint is particularly emphasized in Refs.~\cite{Oda:2021tiv,Oda:2023qek}).

On $\mc H$, we define the position and momentum operators:\footnote{
See Ref.~\cite{Oda:2021tiv} for a detailed discussion on the position operator in the relativistic one-particle subspace.
}
\al{
\op x\Ket{x}
	&=	x\Ket{x},&
\op p\Ket{p}
	&=	p\Ket{p}.
}
It follows that
\al{
\Bra{x}\op p
	&=	-i{\p\ov\p x}\Bra{x},&
\op p\Ket{x}
	&=	\Ket{x}i{\overleftarrow\p\ov\p x},&
\Bra{p}\op x
	&=	i{\p\ov\p p}\Bra{p},&
\op x\Ket{p}
	&=	\Ket{p}\pn{-i{\overleftarrow\p\ov\p p}},
	\label{derivative representation}
}
and
\al{
\commutator{\op x}{\op p}
	&=	i\id,&
\commutator{\op x}{\op x}
	=	\commutator{\op p}{\op p}
	=	0.
}

The projective measurement for position $\M_\tx{pos}$ and that for momentum $\M_\tx{mom}$ are defined by, $\forall\st\rh\in\stf Z{\mc H}$,
\al{
\pdf{\sqbr{M_\tx{pos}\rh}}{x}
	&=	\Tr\Fnl{\Ket{x}\Bra{x}\st\rh},&
\pdf{\sqbr{M_\tx{mom}\rh}}{p}
	&=	\Tr\Fnl{\Ket{p}\Bra{p}\st\rh}.
		\label{measurement of position and momentum}
}
Especially when $\st\rh$ is a pure state $\st\rh=\Ket\psi\Bra\psi$, we recover the familiar PDFs:
\al{
\pdf{\Sqbr{M_\tx{pos}\Ket\psi\Bra\psi}}{x}
	&=	\ab{\Braket{x|\psi}}^2,&
\pdf{\Sqbr{M_\tx{mom}\Ket\psi\Bra\psi}}{p}
	&=	\ab{\Braket{p|\psi}}^2.
}

\subsection{Gaussian basis}
We define the Gaussian wave-packet state $\Ket{ X, P;\sigma}$ on the momentum-space basis~\cite{Ishikawa:2005zc,Ishikawa:2018koj,Ishikawa:2020hph} (see also Ref.~\cite{Oda:2021tiv} for a historical account):
\al{
\Braket{ p| X, P;\sigma}
	&:=	\pn{\sigma\ov\pi}^{1\ov4}\exp\fn{-i pX-{\sigma\ov2}\pn{ p- P}^2},
		\label{Gaussian packet defined}
}
where we have normalized such that
\al{
\Abb{\Ket{ X, P;\sigma}}^2=1.
	\label{normalization of Gaussian}
}
Physically, $X$ and $P$ give the centers of the wave packet in the position and momentum spaces, respectively, and $\sigma$ ($\sigma^{-1}$) roughly gives the spatial (momentum) width-squared.
The corresponding wave function becomes
\al{
\Braket{ x| X, P;\sigma}
	&=	\int_\R\df p\Braket{ x| p}\Braket{ p| X, P;\sigma}
	=	{1\ov\pn{\pi\sigma}^{1\ov4}}
		\exp\fn{i P\pn{ x- X}-{\pn{ x- X}^2\ov2\sigma}}.
		\label{Gaussian wave function 1D}
}

It is important that the Gaussian wave-packet states form a complete basis that spans~$\mc H$: For any fixed $\sigma$,
\al{
\int_{\mathbb R^2}{\df X\,\df P\ov2\pi}\Ket{X,P;\sigma}\Bra{X,P;\sigma}
	&=	\id,
		\label{Gaussian completeness}
}
which can be verified by sandwiching both-hand sides by $\Bra{p}$ and $\Ket{p'}$ (or equally by $\Bra{x}$ and $\Ket{x'}$).\footnote{
Here, it is meant that $\int_{\mathbb R^2}{\df X\,\df P\ov2\pi}:={1\ov2\pi}\int_{-\infty}^\infty\df X\int_{-\infty}^\infty\df P$; see footnote~\ref{integral notation}.
}
It is somewhat beautiful that the phase space measure ${\df X\,\df P\ov2\pi}$ emerges automatically from the normalization~\eqref{normalization of Gaussian}.
Note that the spatial width-squared $\sigma$ is fixed and is not summed in the completeness~\eqref{Gaussian completeness}.

It is also noteworthy that the Gaussian basis states are not mutually orthogonal:
\al{
\Braket{ X, P;\sigma|X',P';\sigma}
	&=	\int_\R\df p\Braket{ X, P;\sigma|p}\Braket{p|X',P';\sigma}\nn
	&=
		\exp\fn{
			i{P+P'\ov2}\pn{ X-X'}
			-{\pn{ X- X'}^2\ov4\sigma}
			-{\sigma\ov4}\pn{ P-P'}^2
				}.
				\label{non-orthogonality}
}
In this sense, the relation~\eqref{Gaussian completeness} is sometimes called the overcompleteness.
The relation~\eqref{Gaussian completeness} tells that the Gaussian basis provides a non-projective POVM for any given fixed $\sigma$:
\al{
\Set{{\Ket{X,P;\sigma}\Bra{X,P;\sigma}\ov2\pi}}_{\pn{X,P}\in\R^2}.
	\label{Gaussian POVM}
}
Finally, the completeness~\eqref{Gaussian completeness} implies that, on $\mc H$,
\al{
\Tr\fnl{\cdots}=\int{\df X\,\df P\ov2\pi}\Bra{X,P;\sigma}\cdots\Ket{X,P;\sigma}.
	\label{trace over Gaussian basis}
}

\section{Separate measurement of position or momentum}\label{separate measurement section}
For the rest of this paper (except in Eq.~\eqref{measurement onto phase space}), we will exclusively focus on measurements of an initial Gaussian pure state:\footnote{
We will see that, under the trace~\eqref{trace over Gaussian basis}, this normalization of the density operator gives the correct probability~\eqref{overlap of packets 1D} and the normalization~\eqref{rhoin properly normalized}.
}
\al{
\st\rhoin
	&=	\Ket{\Xin,\Pin;\sigmain}\Bra{\Xin,\Pin;\sigmain}.
	\label{initial state}
}

In this section, we \emph{separately} measure either its position by $\M_\tx{pos}$ or its momentum by $\M_\tx{mom}$, as defined in Eq.~\eqref{measurement of position and momentum}.
Though we list both the results of position and momentum measurements hereafter for conciseness, it should be kept in mind that either one of the measurements is performed, independently from the other.

Given the initial state~\eqref{initial state}, its measured PDFs in the position and momentum spaces are
\al{
\pdf{\sqbr{M_\tx{pos}\rhoin}}{x}
	&=	\ab{\Braket{ x|\Xin,\Pin;\sigmain}}^2
	=	{1\ov\sqrt{\pi\sigmain}}
		\exp\fn{-{\pn{ x-\Xin}^2\ov\sigmain}},
			\label{Gaussian PDF 1D}\\
\pdf{\sqbr{M_\tx{mom}\rhoin}}{p}
	&=	\ab{\Braket{ p|\Xin,\Pin;\sigmain}}^2
	=	\sqrt{\sigmain\ov\pi}
		\exp\fn{-\sigmain\pn{p-\Pin}^2}.
			\label{Gaussian PDF in momentum space 1D}
}
They are centered around $x=\Xin$ and $ p= \Pin$, with the widths-squared roughly $\sigmain$ and $\sigmain^{-1}$, respectively.

\subsection{Expectation and variance for separate measurements}
We obtain the expectation value, the squared semi-norm, and the variance (given in Eqs.~\eqref{expectation value}, \eqref{quantum standard deviation}, \eqref{classical expectation value}, and \eqref{classical standard deviation}) for $\M_\tx{pos}$ of $\op x$ (left) and for $\M_\tx{mom}$ of $\op p$ (right):
\al{
\Braket{\op x}_{\st\rhoin}
	&=	\Braket{\idf}_\pd{M_\tx{pos}\rhoin}
	=	\Xin,&
\Braket{\op p}_{\st\rhoin}
	&=	\Braket{\idf}_{\pd{M_\tx{mom}\rhoin}}
	=	\Pin,
		\label{expectation values of position and momentum}\\
\abb{\op x}^2_{\st\rhoin}
	&=	\abb{\idf}^2_\pd{M_\tx{pos}\rhoin}
	=	\Xin^2+{\sigmain\ov2},&
\abb{\op p}^2_{\st\rhoin}
	&=	\abb{\idf}^2_\pd{M_\tx{mom}\rhoin}
	=	\Pin^2+{1\ov2\sigmain},
		\label{squared semi-norm}\\
\sigma^2_{\st\rhoin}\fnl{\op x}
	&=	\sigma^2_\pd{M_\tx{pos}\rhoin}\fnl{\idf}
	=	{\sigmain\ov2},&
\sigma^2_{\st\rhoin}\fnl{\op p}
	&=	\sigma^2_\pd{M_\tx{mom}\rhoin}\fnl{\idf}
	=	{1\ov2\sigmain},
}
where $\idf$ is the identity function
\al{
\idff{x}
	&=	x,&
\idff{p}
	&=	p;
	\label{concerete realization of id}
}
recall that $\abb{\op x}^2_{\st\rhoin}:=\Braket{\op x[2]}_{\st\rhoin}$, $\abb{\idf}^2_\pd{M_\tx{pos}\rhoin}:=\Braket{\idf[2]}_\pd{M_\tx{pos}\rhoin}$, etc.

We also have the following results, which are independent of measurements:
\al{
\Braket{\commutator{\op x}{\op p}\ov2i}_{\st\rhoin}
	&=	{1\ov2},&
\Braket{\anticommutator{\op x}{\op p}\ov2}_{\st{\rhoin}}
	&=	\Xin\Pin.
		\label{measurement-independent part}
}
For some readers' ease, we list some matrix elements as well:
\al{
\Braket{ p|\op{x}|\Xin, \Pin;\sigmain}
	&=	\Pn{\Xin-i\sigmain\pn{p-\Pin}}
		\Braket{ p|{\Xin}, \Pin;\sigmain},
			\label{plane-Gaussian matrix element of x}\\
\Braket{ x|\op{p}|\Xin, \Pin;\sigmain}
	&=	\pn{\Pin+i{x-\Xin\ov\sigmain}}
		\Braket{ x|{\Xin}, \Pin;\sigmain},
			\label{plane-Gaussian matrix element of p}
}
which follow from the derivative representation~\eqref{derivative representation} and may be useful to derive expressions below.

\subsection{Pullback and pushforward for separate measurements}
Having completed our review, we now turn to new developments from here on.

Given a function $\rf f\in\rff R\R$, we find its LT adjoint:
\al{
\op{\la M_\tx{pos}f}
	&=	\int_\R\df x\,\rff f{ x}\Ket{ x}\Bra{ x},&
\op{\la M_\tx{mom}f}
	&=	\int_\R\df p\,\rff f{ p}\Ket{ p}\Bra{ p},
		\label{LT adjoint of position and momentum}
}
It is straightforward to check that they satisfy, $\forall\st\rh\in\stf Z{\mc H}$,
\al{
\Braket{\op{\la M_\tx{pos}f}}_{\st\rh}
	&=	\Braket{\rf f}_{\pd{M_\tx{pos}\rh}},&
\Braket{\op{\la M_\tx{mom}f}}_{\st\rh}
	&=	\Braket{\rf f}_{\pd{M_\tx{mom}\rh}},
}
by expanding the general state as
\al{
\st\rh
	&=	\int_\R\df x\int_\R\df x'\,\stf{\rho}{x,x'}\Ket{x}\Bra{x'}
	=	\int_\R\df p\int_\R\df p'\,\stf{\wt\rho}{ p, p'}\Ket{ p}\Bra{ p'},
		\label{decomposition of rho}
}
where $\stf{\rho}{x,x'}:=\Bra{x}\st\rh\Ket{x'}$, with $\stf\rho[*]{x,x'}=\stf\rho{x',x}$ and $\int_\R\df x\,\stf\rho{x,x}=1$ (in particular, it follows that $\pdf{\sqbr{M_\tx{pos}\rh}}x=\stf\rho{x,x}$); and similarly for the momentum space.
The result~\eqref{LT adjoint of position and momentum} is indeed trivial because of the generality~\eqref{projective LT adjoint}, but we spelled it out anyway for concreteness.

Equipped with the definite representation~\eqref{Gaussian packet defined}, yielding the matrix elements~\eqref{plane-Gaussian matrix element of x} and \eqref{plane-Gaussian matrix element of p}, we can explicitly compute the pushforward of $\op x$,
\al{
\rff{\sqbr{M_{\tx{pos}\star}\opa{x}}}{ x}
	&=	x,&
\rff{\sqbr{M_{\tx{mom}\star}\opa{x}}}{ p}
	&=	\Xin,
}
and the pushforward of $\op p$,
\al{
\rff{\sqbr{M_{\tx{pos}\star}\opa{p}}}{ x}
	&=	\Pin,&
\rff{\sqbr{M_{\tx{mom}\star}\opa{p}}}{ p}
	&=	p.
	\label{pushforward for p}
}
One may verify these pushforwards to satisfy the defining relation~\eqref{definition of pushforward}, or more concretely~\eqref{concrete definition of pushforward}:
As an illustration, we show the derivation of $\rf{M_{\tx{pos}\star}\opa{p}}$ given in the left of Eq.~\eqref{pushforward for p}.
All other computations in this paper can be performed quite similarly.
The defining relation~\eqref{definition of pushforward} reads, $\forall\rf f\in\rff R\R$,
\al{
\Braket{\op p,\,\op{\la M_\tx{pos} f}}_{\st{\rhoin}}
	=	\Braket{\rf{M_{\tx{pos}\star}\wh p},\,\rf f}_{\pd{M_\tx{pos}\rhoin}},
}
or more concretely from Eq.~\eqref{concrete definition of pushforward}, $\forall\rf f\in\rff R\R$,
\al{
\Tr\fnl{{\anticommutator{\op p}{\op{\la M_\tx{pos}f}}\ov2}\st\rhoin}
&=
\int_\R\df x\,
	\rff{\sqbr{M_{\tx{pos}\star}\wh p}}x\,
	\rff fx\,
	\pdf{\sqbr{M_\tx{pos}\rhoin}}x.
		\label{pushforward example}
}
Putting the pullback~\eqref{LT adjoint of position and momentum} into the left-hand side, we obtain
\al{
\tx{l.h.s.}
	&=	
		\int_\R\df x\,\rff f{ x}
		\Tr\fnl{
			{\op{p}\Ket{ x}\Bra{ x}+\Ket{ x}\Bra{ x}\op{p}\ov2}
			\Ket{{\Xin}, \Pin;\sigmain}
			\Bra{{\Xin}, \Pin;\sigmain}
			}\nn
	&=	\int_\R\df x\,\rff f{ x}\,\Re\Bigg(
			\Braket{{\Xin},\Pin;\sigmain|x}
			\ub{\Braket{x|\op p|{\Xin},\Pin;\sigmain}}_{
				\pn{\Pin+i{x-\Xin\ov\sigmain}}
					\Braket{ x|{\Xin}, \Pin;\sigmain}
				}
			\Bigg)\nn
	&=	\int_\R\df x\,\rff f{ x}\,\Pin\,\ub{\Ab{\Braket{ x|{\Xin}, \Pin;\sigmain}}^2}_{\pdf{\sqbr{M_\tx{pos}\rhoin}}x},
}
where $\Re$ denotes the real part and we used Eq.~\eqref{plane-Gaussian matrix element of p} in the second last line.
Comparing with the right-hand side of Eq.~\eqref{pushforward example}, we obtain the constant solution $\rff{\sqbr{M_{\tx{pos}\star}\opa{p}}}{ x}=\Pin$.

\subsection{LT error for separate measurements}
Having obtained the pushforward, we can now compute the classical semi-norm for $\op x$,
\al{
\abb{\rf{M_{\tx{pos}\star}\wh x}}^2_\pd{M_\tx{pos}\rhoin}
	&=	\Xin^2+{\sigmain\ov2},&
\abb{\rf{M_{\tx{mom}\star}\wh x}}^2_{\pd{M_\tx{pos}\rhoin}}
	&=	\Xin^2,
}
and the classical semi-norm for $\op p$,
\al{
\abb{\rf{M_{\tx{pos}\star}\wh p}}^2_{\pd{M_{mom}\rhoin}}
	&=	\Pin^2,&
\abb{\rf{M_{\tx{mom}\star}\wh p}}^2_{\pd{M_{mom}\rhoin}}
	=	\Pin^2+{1\ov2\sigmain}.
}
Combining with the quantum semi-norm~\eqref{squared semi-norm}, the resultant LT error~\eqref{LT error defined} is, for $\op x$,
\al{
\varepsilon^2_{\st{\rhoin}}\fnl{\op{x};\M_\tx{pos}}
	&=	0,&
\varepsilon^2_{\st{\rhoin}}\fnl{\op{x};\M_\tx{mom}}
	&=	{\sigmain\ov2},
}
and, for $\op p$,
\al{
\varepsilon^2_{\st{\rhoin}}\fnl{\op{p};\M_\tx{pos}}
	&=	{1\ov2\sigmain},&
\varepsilon^2_{\st{\rhoin}}\fnl{\op{p};\M_\tx{mom}}
	&=	0.
}

We see that for both the position and momentum measurements, the left-hand side of the LT inequality~\eqref{LT inequality} vanishes:
\al{
\varepsilon_{\st{\rhoin}}\fnl{\op{x};\M_\tx{pos}}
\varepsilon_{\st{\rhoin}}\fnl{\op{p};\M_\tx{pos}}
	&=	0,&
\varepsilon_{\st{\rhoin}}\fnl{\op{x};\M_\tx{mom}}
\varepsilon_{\st{\rhoin}}\fnl{\op{p};\M_\tx{mom}}
	&=	0.
}
Let us see if the right-hand side vanishes consistently.

\subsection{LT inequality for separate measurements}
The pullback of pushforward is, for $\op x$,
\al{
\op{\la M_\tx{pos}M_{\tx{pos}\star}\opa x}
	&=	\op x,&
\op{\la M_\tx{mom}M_{\tx{mom}\star}\opa x}
	&=	\Xin\id,
}
and for $\op p$,
\al{
\op{\la M_\tx{pos}M_{\tx{pos}\star}\opa p}
	&=	\Pin\id,&
\op{\la M_\tx{mom}M_{\tx{mom}\star}\opa p}
	&=	\op p.
}
Therefore for the position measurement,
\al{
\commutator{\op{\la M_\tx{pos}M_{\tx{pos}\star}\opa x}}{\op p}
	&=	i\id,&
\commutator{\op x}{\op{\la M_\tx{pos}M_{\tx{pos}\star}\opa p}}
	&=	0,
}
and for the momentum measurement,
\al{
\commutator{\op{\la M_\tx{mom}M_{\tx{mom}\star}\opa x}}{\op p}
	&=	0,&
\commutator{\op x}{\op{\la M_\tx{mom}M_{\tx{mom}\star}\opa p}}
	&=	i\id.
}
Combining with the measurement-independent part~\eqref{measurement-independent part}, we see that the quantum contribution to the LT lower bound~\eqref{imaginary part for LT bound} vanishes:
\al{
\mc I_{\st\rhoin}\fnl{\op x,\op p;\M_\tx{pos}}
	&=	0,&
\mc I_{\st\rhoin}\fnl{\op x,\op p;\M_\tx{mom}}
	&=	0.
}

We can compute
\al{
\Braket{\rf{M_{\tx{pos}\star}\wh x},\,\rf{M_{\tx{pos}\star}\wh p}}_\pd{M_\tx{pos}\rhoin}
	&=	\Braket{\idf}_\pd{M_\tx{pos}\rhoin}\Pin
	=	\Xin\Pin,\\
\Braket{\rf{M_{\tx{mom}\star}\wh x},\,\rf{M_{\tx{mom}\star}\wh p}}_\pd{M_\tx{mom}\rhoin}
	&=	\Xin\Braket{\idf}_\pd{M_\tx{mom}\rhoin}
	=	\Xin\Pin,
}
where we used Eq.~\eqref{expectation values of position and momentum}.
Combining with the measurement-independent part~\eqref{measurement-independent part}, we see that the semi-classical contribution to the LT lower bound~\eqref{real part for LT bound} also vanishes:
\al{
\mc R_{\st\rhoin}\fnl{\op x,\op p;\M_\tx{pos}}
	&=	0,&
\mc R_{\st\rhoin}\fnl{\op x,\op p;\M_\tx{mom}}
	&=	0.
}
The LT inequality turned out to be $0=0$!

\subsection{Lee error for separate measurement?}
To compute the Lee error~\eqref{Lee error}, we need the inverse of the pullback~\eqref{LT adjoint of position and momentum}. It is trivial for (so to say) the ``diagonal'' measurement:
\al{
\rff{\sqbr{M_\tx{pos}^{\star-1}\opa x}}x
	&=	x,&
\rff{\sqbr{M_\tx{mom}^{\star-1}\opa p}}p
	&=	p.
}
It is straightforward to check that they satisfy, $\forall\st\rh\in\stf Z{\mc H}$,
\al{
\Braket{\op x}_{\st\rh}
	&=	\Braket{\rf{M_\tx{pos}^{\star-1}\opa x}}_{\pd{M_\tx{pos}\rh}},&
\Braket{\op p}_{\st\rh}
	&=	\Braket{\rf{M_\tx{mom}^{\star-1}\opa p}}_{\pd{M_\tx{mom}\rh}},
}
by expanding the general state as in Eq.~\eqref{decomposition of rho}.

However, it is hard to obtain the inverse of pullback for the ``off-diagonal'' measurements $\rff{\sqbr{M_\tx{mom}^{\star-1}\opa x}}p$ and $\rff{\sqbr{M_\tx{pos}^{\star-1}\opa p}}x$. For example for the latter, we need to find $\rf f\in\rff R\R$ that satisfies $
\op p
	=
		\int_\R\df x\,\rff fx\,\Ket x\Bra x$ to match Eq.~\eqref{LT adjoint of position and momentum}.
However, the derivative representation~\eqref{derivative representation} leads to
\al{
\op p
	&=	\int_\R\df x\int_\R\df x'\Ket x\Braket{x|\op p|x'}\Bra{x'}
	=	\int_\R\df x\int_\R\df x'\pn{-i{\p\ov\p x}\delta\fn{x-x'}}\Ket x\Bra{x'}.
}
Formally, the solution (if it exists) can be expressed as $\rff fx=\int_\R\df x'\pn{-i{\p\ov\p x}\delta\fn{x-x'}}$.\footnote{
We can find $\rf f\in\rff R\R$ that satisfies $\rff fx\Bra x=\int_\R\df x'\pn{-i{\p\ov\p x}\delta\fn{x-x'}}\Bra{x'}$ as follows: $\Longleftrightarrow
\rff fx\delta\fn{x-x''}=\int_\R\df x'\pn{-i{\p\ov\p x}\delta\fn{x-x'}}\delta\fn{x'-x''}=\pn{-i{\p\ov\p x}\delta\fn{x-x''}}\Longleftrightarrow\rff fx=\int_\R\df x'\pn{-i{\p\ov\p x}\delta\fn{x-x'}}$.
}
This expression evidently requires regularization due to the derivative of the Dirac delta function. For example, using the regularization $\delta_\sigma\fn{x-x'}=\pn{\pi\sigma}^{-1/2}\exp\Fn{-\pn{x-x'}^2/\sigma}$ leads to a vanishing result for $\rff fx$, indicating the need for further elaboration. Below, we will demonstrate that a joint measurement in the Gaussian phase space yields a naturally regularized expression, as will be detailed in Eq.~\eqref{decomposition of position and momentum}.

\section{Joint measurement in phase space}\label{Gaussian joint measurement section}
Below, we first review known facts about the Gaussian matrix element, then restart with our main new development from Sec.~\ref{phase space projection section}.

The completeness~\eqref{Gaussian completeness} implies that the Gaussian basis can expand \emph{any} function, including another Gaussian wave packet with a different width-squared:
\al{
\Ket{\Xin,\Pin;\sigmain}
	&=	\int{\df X\df P\ov2\pi}\Ket{X,P;\sigma}\Braket{X,P;\sigma|\Xin,\Pin;\sigmain}.
}
The expansion coefficient can be computed as~\cite{Ishikawa:2020hph}
\al{
\Braket{ X, P;\sigma|\Xin,\Pin;\sigmain}
	&=	\int_\R\df p\Braket{ X, P;\sigma|p}\Braket{p|\Xin,\Pin;\sigmain}\nn
	&=
		\sqrt2\pn{\sigmared\ov\sigmasum}^{1\ov4}
		\exp\fn{
			i\ol{ P}\pn{ X-\Xin}
			-{\pn{ X- \Xin}^2\ov2\sigmasum}
			-{\sigmared\ov2}\pn{ P-\Pin}^2
				},
				\label{overlap 1D}
}
where\footnote{
We call the harmonic mean $\sigmared$ the reduced width-squared, in analogy to reduced mass.
}
\al{
\sigmasum
	&:=	\sigma+\sigmain,&
\sigmared
	&:=	{1\ov{1\ov\sigma}+{1\ov\sigmain}}
	=	{\sigma\sigmain\ov\sigma+\sigmain}.
}

Physically, the inner product~\eqref{overlap 1D} can be understood as the transition amplitude from the initial state $\Ket{\Xin,\Pin;\sigmain}$ to the final state $\Ket{X,P;\sigma}$. 
We may find the corresponding transition probability as follows. (See also Appendix~\ref{Ozawa section} for more realistic modeling of measurement processes.)
When integrated all over the phase space, the normalization~\eqref{normalization of Gaussian} and the resultant completeness~\eqref{Gaussian completeness} provide
\al{
\int_{\R^2}{\df X\,\df P\ov2\pi}\ab{\Braket{ X, P;\sigma| \Xin, \Pin;\sigmain}}^2
	&=	1.
		\label{normalization of the Gaussian packet}
}
This gives a natural physical interpretation: Starting from the state $\Ket{\Xin,\Pin;\sigmain}$, the probability $\df\mc P$ of finding a result in the phase space volume $\df X\,\df P\ov2\pi$ around the phase-space point $\pn{X,P}$ is
\al{
\df\mc P\fn{X,P}
	&=	\ab{\Braket{ X, P;\sigma| \Xin, \Pin;\sigmain}}^2{\df X\,\df P\ov2\pi},
}
and in particular, the probability of finding \emph{exactly} the state $\Ket{X,P;\sigma}$ right at the single phase-space point $\pn{X,P}$ is
\al{
\mc P\fn{X,P}
	&=	\int\df\mc P\fn{X',P'}\,\delta\fn{X-X'}\delta\fn{P-P'}\nn
	&=	{\ab{\Braket{ X, P;\sigma| \Xin, \Pin;\sigmain}}^2\ov2\pi}
	=	
		\pn{{1\ov\sqrt{\pi\sigmasum}}e^{-{1\ov\sigmasum}\pn{X-\Xin}^2}}
		\pn{\sqrt{\sigmared\ov\pi}e^{-\sigmared\pn{P-\Pin}^2}}.
			\label{overlap of packets 1D}
}
We see that the detection probability of the particle is centered around
$X=\Xin$ and $P=\Pin$, with the width-squared roughly $\sigmasum$ and $\sigmared^{-1}$, respectively.\footnote{
It is somewhat amusing that even if we provide the POVM $\Set{\Ket{X,P;\sigmain}\Bra{X,P;\sigmain}/2\pi}_{\pn{X,P}\in\R^2}$, the probability of finding the original state $\Ket{\Xin,\Pin;\sigmain}$ is $1/2\pi \simeq 16\% < 1$ due to the non-orthogonality of the basis.
}

From Eq.~\eqref{normalization of the Gaussian packet}, we can check that the density operator~\eqref{initial state} is properly normalized under the trace~\eqref{trace over Gaussian basis}:\footnote{
Note that the cyclic property of trace does not apply at the bra-ket level:
$\Tr\Fnl{\Ket{\psi}\Bra{\psi}}\neq\Tr\Fnl{\Braket{\psi|\psi}}=\Braket{\psi|\psi}\Tr\Fnl{\op1}\neq\Braket{\psi|\psi}$.
}
\al{
\Tr\fnl{\st\rhoin}
	&=	1.
	\label{rhoin properly normalized}
}
In particular, the expectation value of an operator $\op A$ becomes
\al{
\Braket{\op A}_{\st\rhoin}
	=	\Tr\fnl{\op A\st\rhoin}
	&=	\Bra{\Xin,\Pin;\sigmain}\op A\Ket{\Xin,\Pin;\sigmain}.
}

When we do not observe the momentum, we integrate the probability~\eqref{overlap of packets 1D} over all the possible momenta:
\al{
\int_\R\df P
{\ab{\Braket{ X, P;\sigma| \Xin, \Pin;\sigmain}}^2\ov2\pi}
	&=	{1\ov\sqrt{\pi\sigmasum}}\exp\fn{-{\pn{X-\Xin}^2\ov\sigmasum}}.
}
In the limit of detector spatial resolution finer than the initial wave packet size, $\sigma\ll\sigmain$, we obtain $\sigmasum\to\sigmain$, recovering the position-space PDF: $\pdf{\sqbr{M_\tx{pos}\rhoin}}{X}=\ab{\Braket{X|\Xin,\Pin;\sigmain}}^2$; see Eq.~\eqref{Gaussian PDF 1D}.
In the opposite limit $\sigma\gg\sigmain$, the probability is governed by the detector spatial resolution, ${1\ov\sqrt{\pi\sigma}}\exp(-{1\ov\sigma}\pn{X-\Xin}^2)$, which reduces to $\delta\fn{X-\Xin}$ in the limit $\sigma\to0$ (while keeping $\sigma\gg\sigmain$), whereas in the opposite (physically more reasonable) limit $\sigma\to\infty$, the position dispersion becomes infinitely large, and the information of $\Xin$ is also lost.

When we do not observe the position, we integrate the probability~\eqref{overlap of packets 1D} over all the possible positions:
\al{
\int_\R\df X
{\ab{\Braket{ X, P;\sigma| \Xin, \Pin;\sigmain}}^2\ov2\pi}
	&=	\sqrt{\sigmared\ov\pi}\exp\fn{-\sigmared\pn{P-\Pin}^2}.
}
In the limit of detector spatial resolution coarser than the initial wave packet size, $\sigma\gg\sigmain$, we obtain $\sigmared\to\sigmain$, recovering the momentum-space PDF: $\pdf{\sqbr{M_\tx{mom}\rhoin}}{P}=\ab{\Braket{P|\Xin,\Pin;\sigmain}}^2$; see Eq.~\eqref{Gaussian PDF in momentum space 1D}.
In the opposite limit $\sigma\ll\sigmain$, the probability is governed by the detector resolution, $\sqrt{\sigma\ov\pi}\exp(-\sigma\pn{P-\Pin}^2)$, which reduces to $\delta\fn{P-\Pin}$ in the limit $\sigma\to\infty$ (while keeping $\sigma\ll\sigmain$), whereas in the opposite (physically more reasonable) limit $\sigma\to0$, the momentum dispersion becomes infinitely large, and the information of $\Pin$ is also lost.

For reader's ease, we list some matrix elements, which may be useful to derive expressions below: Using the derivative representation~\eqref{derivative representation}, we obtain
\al{
\Braket{X,P;\sigma|\op x|\Xin,\Pin,\sigmain}
	&=	\pn{\ol X-i\sigmared\pn{P-P_\tx{in}}}\Braket{X,P;\sigma|\Xin,\Pin,\sigmain},
		\label{matrix element of x}\\
\Braket{X,P;\sigma|\op p|\Xin,\Pin,\sigmain}
	&=	\pn{\ol P+i{X-X_\tx{in}\ov\sigmasum}}\Braket{X,P;\sigma|\Xin,\Pin;\sigmain},
		\label{matrix element of p}
}
where
\al{
\ol X
	&:=	\sigmared\pn{{X\ov\sigma}+{X_\tx{in}\ov\sigma_\tx{in}}},&
\ol P
	&:=	{\sigma P+\sigma_\tx{in}P_\tx{in}\ov\sigma+\sigma_\tx{in}}.
	\label{X bar P bar}
}

\subsection{Projective measurement onto phase space}\label{phase space projection section}
We assert that the transition probability~\eqref{overlap of packets 1D} can be regarded in the LT formalism as the result of the \emph{POVM measurement onto  phase space}~$\M_\tx{ph}$ defined by, $\forall\st\rh\in\stf Z{\mc H}$,\footnote{\label{projective footnote}
This measurement $\M_\tx{ph}$ is non-projective since the states $\Set{\Ket{X,P;\sigma}}_{\pn{X,P}\in\mathbb R^2}$ are not mutually orthogonal, as demonstrated in Eq.~\eqref{non-orthogonality}.
}
\al{
\pdf{\sqbr{M_\tx{ph}\rh}}{X,P}
	&=	\Tr\fnl{{\Ket{X,P;\sigma}\Bra{X,P;\sigma}\ov2\pi}\st\rh},
	\label{measurement onto phase space}
}
which is nothing but the Husimi Q-function~\cite{1940264}.\footnote{
We thank Gen Kimura for pointing out this connection.
}

Given an initial state $\Ket{\Xin,\Pin;\sigmain}$, or more precisely the initial pure state~\eqref{initial state}, we recover the PDF~\eqref{overlap of packets 1D}:
\al{
\pdf{\sqbr{M_\tx{ph}\rhoin}}{X,P}
	&=	{\ab{\Braket{ X, P;\sigma| \Xin, \Pin;\sigmain}}^2\ov2\pi}.
		\label{joint PDF}
}
For some readers' ease, we list some results of the integrals:
\al{
\Braket{\rf X}_\pd{M_\tx{ph}\rhoin}
	&=	\Xin,&
\Braket{\rf P}_\pd{M_\tx{ph}\rhoin}
	&=	\Pin,&
\Braket{\rf X\rf P}_\pd{M_\tx{ph}\rhoin}
	&=	\Xin\Pin,\nn
\Braket{\rf{\ol X}}_\pd{M_\tx{ph}\rhoin}
	&=	\Xin,&
\Braket{\rf{\ol P}}_\pd{M_\tx{ph}\rhoin}
	&=	\Pin,&
\Braket{\rf{\ol X}\,\rf{\ol P}}_\pd{M_\tx{ph}\rhoin}
	&=	\Xin\Pin,\nn
\Braket{\rf{X^2}}_\pd{M_\tx{ph}\rhoin}
	&=	\Xin^2+{\sigmasum\ov2},&
\Braket{\rf{P^2}}_\pd{M_\tx{ph}\rhoin}
	&=	\Pin^2+{1\ov2\sigmared},\nn
\Braket{\rf{\ol X^2}}_\pd{M_\tx{ph}\rhoin}
	&=	\Xin^2+{\sigmain^2\ov2\sigmasum},&
\Braket{\rf{\ol P^2}}_\pd{M_\tx{ph}\rhoin}
	&=	\Pin^2+{\sigma\ov2\sigmain\sigmasum},
		\label{integral results}
}
where $\rff X{X,P}:=X$ and $\rff P{X,P}:=P$ as well as $\rff{\ol X}{X,P}:=\ol X$ and $\rff{\ol P}{X,P}:=\ol P$, with $\ol X$ and $\ol P$ being given in Eq.~\eqref{X bar P bar}, respectively.

Now we consider the marginalization~\eqref{marginalization}.
We define the classical projection (marginalization) processes $\bm\pi_\tx{pos}$ and $\bm\pi_\tx{mom}$ by, $\forall\pd p\in\pdf W{\R^2}$,
\al{
\pdf{\sqbr{\pi_\tx{pos}p}}{X}
	&:=	\int_\R\df P\,\pdf p{X,P},&
\pdf{\sqbr{\pi_\tx{mom}p}}{P}
	&:=	\int_\R\df X\,\pdf p{X,P},
}
with their trivial LT adjoints~\eqref{adjoint of projection}: $\forall\rf f\in\rff R\R$,
\al{
\rff{\sqbr{\la \pi_\tx{pos}f}}{X,P}
	&=	\rf f\fn{X},&
\rff{\sqbr{\la \pi_\tx{mom}f}}{X,P}
	&=	\rf f\fn{P}.
}
More concretely,
\al{
\pdf{\sqbr{\pi_\tx{pos}M_\tx{ph}\rhoin}}{X}
	&=	\int_\R\df P{\ab{\Braket{ X, P;\sigma| \Xin, \Pin;\sigmain}}^2\ov2\pi}
	=	{1\ov\sqrt{\pi\sigmasum}}
		\exp\fn{-{\pn{ X- \Xin}^2\ov\sigmasum}},
			\label{position distribution from phase space}
			\\
\pdf{\sqbr{\pi_\tx{mom}M_\tx{ph}\rhoin}}{P}
	&=	\int_\R\df X{\ab{\Braket{ X, P;\sigma| \Xin, \Pin;\sigmain}}^2\ov2\pi}
	=	\sqrt{\sigmared\ov\pi}\exp\fn{-\sigmared\pn{ P- \Pin}^2}.
	\label{momentum distribution from phase space}
}
For the Gaussian pure state~\eqref{initial state}, the joint PDF~\eqref{joint PDF} has become separable:
\al{
\pdf{\sqbr{M_\tx{ph}\rhoin}}{X,P}
	&=	
		\pdf{\sqbr{\pi_\tx{pos}M_\tx{ph}\rhoin}}{X}\,
		\pdf{\sqbr{\pi_\tx{mom}M_\tx{ph}\rhoin}}{P}.
}

We find that, in Eqs.~\eqref{position distribution from phase space} and \eqref{momentum distribution from phase space}, the dispersions of position and momentum are $\sigma_\tx{sum}$ and $\sigma_\tx{red}$, respectively.
 These are physically reasonable because in the limit of large spatial width of either the initial packet or the detector spatial resolution ($\sigmain\to\infty$ or $\sigma\to\infty$), the spatial dispersion $\sigma_\tx{sum}$ goes to infinity, while the momentum dispersion $\sigma_\tx{red}$ remains $\sigma$ or $\sigmain$ in the respective cases. The parallel argument holds in the limit of zero spatial width ($\sigmain\to0$ or $\sigma\to0$) for either the initial packet or the detector spatial resolution.

We see that the Gaussian measurement smoothly interpolates the two limiting measurements in the momentum space and in the position space:
\begin{itemize}
\item In the limit of detector spatial resolution finer than the original wave-packet size, $\sigma\ll\sigmain$, we obtain $\sigmasum\to\sigmain$ and $\sigmared\to\sigma$, and hence
\al{
\pdf{\sqbr{\pi_\tx{pos}M_\tx{ph}\rhoin}}{X}
	&\to	{1\ov\sqrt{\pi\sigmain}}
			\exp\fn{-{\pn{ X- \Xin}^2\ov\sigmain}}
	=	\pdf{\sqbr{M_\tx{pos}\rhoin}}{X},\\
\pdf{\sqbr{\pi_\tx{mom}M_\tx{ph}\rhoin}}{P}
	&\to	\sqrt{\sigma\ov\pi}\exp\fn{-\sigma\pn{ P- \Pin}^2}.
}
We have recovered the original PDF in the position space: $\ab{\Braket{X|\Xin,\Pin;\sigmain}}^2$ (see Eq.~\eqref{Gaussian PDF 1D}), whereas the information of the original wave-packet size is lost in the momentum space. Further limit of spatially coarse detector $\sigma\to\infty$ (while keeping $\sigma\ll\sigmain$) gives $\pdf{\sqbr{\pi_\tx{mom}M_\tx{ph}\rhoin}}{P}\to\delta\fn{P-\Pin}$, whereas in the opposite (physically more reasonable) limit of spatially fine detector $\sigma\to0$ makes the momentum dispersion infinitely large, leading to the loss of information of $\Pin$.
\item In the limit of detector spatial resolution coarser than the original wave-packet size, $\sigma\gg\sigmain$, we obtain $\sigmasum\to\sigma$ and $\sigmared\to\sigmain$, and hence
\al{
\pdf{\sqbr{\pi_\tx{pos}M_\tx{ph}\rhoin}}{X}
	&\to	{1\ov\sqrt{\pi\sigma}}
		\exp\fn{-{1\ov\sigma}\pn{ X- \Xin}^2},\\
\pdf{\sqbr{\pi_\tx{mom}M_\tx{ph}\rhoin}}{P}
	&\to	\sqrt{\sigmain\ov\pi}\exp\fn{-\sigmain\pn{ P- \Pin}^2}=	\pdf{\sqbr{M_\tx{mom}\rhoin}}{P}.
}
We have recovered the original PDF in the momentum space $\ab{\Braket{P|\Xin,\Pin;\sigmain}}^2$ (see Eq.~\eqref{Gaussian PDF in momentum space 1D}), whereas the information of the original wave-packet size is lost in the position space.
Further limit of spatially fine detector $\sigma\to0$ (while keeping $\sigma\gg\sigmain$) gives $\pdf{\sqbr{\pi_\tx{pos}M_\tx{ph}\rhoin}}{X}\to\delta\fn{X-\Xin}$, whereas in the opposite (physically more reasonable) limit of spatially coarse detector $\sigma\to\infty$ makes the position dispersion infinitely large, leading to the loss of information of $\Xin$.
\end{itemize}

\subsection{Pullback and pushforward for joint measurement}
One might find the LT adjoint $\la\M_\tx{ph}$ slightly less trivial because of the non-orthogonality of the basis~\eqref{non-orthogonality}; see also footnote~\ref{projective footnote}. However, we find that the result is diagonal too: For a given function $\rf f\in\rff R{\R^2}$, its LT adjoint is
\begin{align}
\op{\la M_\tx{ph}f}
    &=  \int_{\R^2}\df X\,\df P\,\rff f{X,P}{\Ket{X,P;\sigma}\Bra{X,P;\sigma}\ov2\pi},
        \label{LT adjoint for joint measurement}
\end{align}
which can be verified by inserting the expansion~\eqref{decomposition of rho} into the defining relation
$
\langle\op{\la M_\tx{ph}f}\rangle_{\st\rh}
    =   \Braket{\rf f}_{\pd{M_\tx{ph}\rh}}$.
This shows that the LT adjoint maintains its diagonal nature even in the context of the POVM measurement~\eqref{measurement onto phase space}.

The position and momentum operators have the following decomposition in the Gaussian phase space:
\al{
\op x
	&=	\int_{\R^2}\df X\,\df P\,X{\Ket{X,P;\sigma}\Bra{X,P;\sigma}\ov2\pi},&
\op p
	&=	\int_{\R^2}\df X\,\df P\,P{\Ket{X,P;\sigma}\Bra{X,P;\sigma}\ov2\pi},
		\label{decomposition of position and momentum}
}
which can be verified by sandwiching with $\Bra{x}$ and $\Ket{x'}$ for $\op x$, and with $\Bra{p}$ and $\Ket{p'}$ for $\op p$. We can also confirm this by sandwiching with $\Bra{X',P';\sigma}$ and $\Ket{X'',P'';\sigma}$, and then using Eq.~\eqref{non-orthogonality}, \eqref{matrix element of x}, and \eqref{matrix element of p}.
Comparing with Eq.~\eqref{LT adjoint for joint measurement} for the general LT adjoint/pullback, we obtain its inverse:
\al{
\rff{\sqbr{M_\tx{ph}^{\star-1}\opa x}}{X,P}
	&=	X,&
\rff{\sqbr{M_\tx{ph}^{\star-1}\opa p}}{X,P}
	&=	P.
}
From Eq.~\eqref{integral results}, we obtain their classical semi-norms,
\al{
\abb{\rf{M_\tx{ph}^{\star-1}\opa x}}^2_\pd{M_\tx{ph}\rhoin}
	&=	\Xin^2+{\sigmasum\ov2},&
\abb{\rf{M_\tx{ph}^{\star-1}\opa p}}^2_\pd{M_\tx{ph}\rhoin}
	&=	\Pin^2+{1\ov2\sigmared},
	\label{classical semi-norms}
}
and their classical semi-inner product,
\al{
\Braket{\rf{M_\tx{ph}^{\star-1}\opa x},\,\rf{M_\tx{ph}^{\star-1}\opa p}}_\pd{M_\tx{ph}\rhoin}
	&=	\Xin\Pin.
		\label{classical semi-inner product}
}

Similarly, we obtain the pushforward of $\op x$ and $\op p$ (over $\st\rh$):
\al{
\rff{\sqbr{M_{\tx{ph}\star}\opa{x}}}{X,P}
	&=	\ol X,&
\rff{\sqbr{M_{\tx{ph}\star}\opa{p}}}{X,P}
	&=	\ol P,
}
where $\ol X$ and $\ol P$ are given in Eq.~\eqref{X bar P bar}.
From Eq.~\eqref{integral results}, we obtain
\al{
\abb{\rf{M_{\tx{ph}\star}\opa x}}^2_\pd{M_\tx{ph}\rhoin}
	&=	\Xin^2+{\sigmain^2\ov2\sigmasum},&
\abb{\rf{M_{\tx{ph}\star}\opa p}}^2_\pd{M_\tx{ph}\rhoin}
	&=	\Pin^2+{\sigma\ov2\sigmain\sigmasum},
		\label{semi-norm squared of pushforward}
}
and
\al{
\Braket{\rf{M_{\tx{ph}\star}\opa x},\,\rf{M_{\tx{ph}\star}\opa p}}_\pd{M_\tx{ph}\rhoin}
	&=	\Xin\Pin.
}

Accordingly, the pullback of pushforward is
\al{
\op{\la M_\tx{ph}M_{\tx{ph}\star}\opa x}
	&=	\int_{\R^2}\df X\,\df P\,\ol X{\Ket{X,P;\sigma}\Bra{X,P;\sigma}\ov2\pi},
		\label{pull push of x}\\
\op{\la M_\tx{ph}M_{\tx{ph}\star}\opa p}
	&=	\int_{\R^2}\df X\,\df P\,\ol P{\Ket{X,P;\sigma}\Bra{X,P;\sigma}\ov2\pi}.
		\label{pull push of p}
}

\subsection{LT error for joint measurement}
Subtracting the classical semi-norm squared of pushforward~\eqref{semi-norm squared of pushforward} from the quantum one of the original operator~\eqref{squared semi-norm}, we obtain the resultant LT errors~\eqref{LT error defined} for $\op x$ and $\op p$:
\al{
\varepsilon^2_{\st{\rhoin}}\fnl{\op{x};\M_\tx{ph}}
	&=	{\sigmared\ov2}, &
\varepsilon^2_{\st{\rhoin}}\fnl{\op{p};\M_\tx{ph}}
	&=	{1\ov2\sigmasum}.
		\label{LT errors for Gaussian initial state}
}
We see that the left-hand side of the LT inequality~\eqref{LT inequality} becomes
\al{
\varepsilon_{\st{\rhoin}}\fnl{\op{x};\M_\tx{ph}}
\varepsilon_{\st{\rhoin}}\fnl{\op{p};\M_\tx{ph}}
	&=	{1\ov2}\sqrt{\sigmared\ov\sigmasum}.
	\label{lhs for LT inequality}
}

In the limits of spatially fine and coarse detectors $\sigma\to0$ and $\infty$, respectively, we obtain
\al{
\varepsilon_{\st{\rhoin}}^2\fnl{\op{x};\M_\tx{ph}}
	&\to\begin{cases}
		0	&(\sigma\to0),\\
		{\sigmain\ov2}	&(\sigma\to\infty),
		\end{cases}
		&
\varepsilon_{\st{\rhoin}}^2\fnl{\op{p};\M_\tx{ph}}
	&\to\begin{cases}
		{1\ov2\sigmain}	&(\sigma\to0),\\
		0	&(\sigma\to\infty).
		\end{cases}
}
The product of LT errors~\eqref{lhs for LT inequality} goes to zero in both the limits $\sigma\to0$ and $\infty$, while it takes the maximum value $1/4$ when $\sigma=\sigma_\tx{in}$.

As discussed in the paragraph following the one containing Eqs.~\eqref{position distribution from phase space} and \eqref{momentum distribution from phase space}, the natural spatial and momentum dispersions are $\sigma_\tx{sum}$ and $\sigma_\tx{red}$, respectively. However, the dispersions appearing in the LT errors~\eqref{LT errors for Gaussian initial state} are opposite to these natural ones. This should reflect the fact that the LT errors correspond to the Ozawa errors, whose product can even become zero. We will return to this point in Sec.~\ref{sec: Physical interpretation for uncertainty relations}.

\subsection{LT inequality for joint measurement}
Using Eqs.~\eqref{pull push of x} and \eqref{pull push of p}, and then Eqs.~\eqref{matrix element of x} and \eqref{matrix element of p}, we obtain
\al{
\Braket{\commutator{\op{\la M_\tx{ph}M_{\tx{ph}\star}\opa x}}{\op p}\ov2i}_{\st\rhoin}
	&=	\Im\Braket{\rf{\ol X}\pn{\rf{\ol P}+i{\rf X-\Xin\ov\sigmasum}}}_\pd{M_\tx{ph}\rhoin}
	=	{\sigmain\ov2\sigmasum},\\
\Braket{\commutator{\op x}{\op{\la M_\tx{ph}M_{\tx{ph}\star}\opa p}}\ov2i}_{\st\rhoin}
	&=	\Im\Braket{\pn{\rf{\ol X}-i\sigmared\pn{\rf P-P_\tx{in}}}^*\rf{\ol P}}_\pd{M_\tx{ph}\rhoin}
	=	{\sigma\ov2\sigmasum},
}
where $\Im$ denotes the imaginary part and we used Eq.~\eqref{integral results} in the last step of each.
Subtracting the above two from the measurement-independent part in the former of Eq.~\eqref{measurement-independent part}, we see that the quantum contribution to the LT lower bound~\eqref{imaginary part for LT bound} vanishes:
\al{
\mc I_{\st\rhoin}\fnl{\op x,\op p;\M_\tx{ph}}
	&=	0.
}

For the semi-classical contribution to the LT lower bound~\eqref{real part for LT bound},
we subtract the measurement-dependent part~\eqref{classical semi-inner product} from the independent part in the latter of Eq.~\eqref{measurement-independent part}. As a result, we find it to vanish:
\al{
\mc R_{\st\rhoin}\fnl{\op x,\op p;\M_\tx{ph}}
	&=	0.
}

It is important that the LT inequality gives no lower bound for the joint measurement of position and momentum (see Sec.~\ref{sec: Physical interpretation for uncertainty relations} for further discussion).
The LT inequality~\eqref{LT inequality} now becomes
\al{
{1\ov2}\sqrt{\sigmared\ov\sigmasum}
	&\geq	0,
		\label{LT inequality for Gaussian initial state}
}
which is saturated in the limits $\sigma\to0$ and $\infty$.
The left-hand side takes the maximum value $1/4$ at $\sigma=\sigmain$.

\subsection{Lee error for joint measurement}
Subtracting the quantum semi-norm~\eqref{squared semi-norm} from the classical~\eqref{classical semi-norms}, we obtain the Lee errors~\eqref{Lee error} for $\op x$ and $\op p$:
\al{
\wt\varepsilon^2_{\st{\rhoin}}\fnl{\op{x};\M_\tx{ph}}
	&=	{\sigma\ov2}, &
\wt\varepsilon^2_{\st{\rhoin}}\fnl{\op{p};\M_\tx{ph}}
	&=	{1\ov2\sigma}.
}
It is remarkable that, unlike the LT error, the Lee error is solely determined from the detector resolution, independent of the width of the measured state.

Their product gives the left-hand side of the Lee inequality:
\al{
\wt\varepsilon_{\st{\rhoin}}\fnl{\op{x};\M_\tx{ph}}
\wt\varepsilon_{\st{\rhoin}}\fnl{\op{p};\M_\tx{ph}}
	&=	{1\ov2}.
	\label{product of Lee errors}
}

\subsection{Lee inequality for joint measurement}
The quantum contribution to the Lee lower bound~\eqref{imaginary for Lee inequality} is identical to that of Heisenberg's, arising from the measurement-independent part, in the former of Eq.~\eqref{measurement-independent part}:
\al{
\mc I_{0\st\rhoin}\fnl{\op x,\op p}
	&=	{1\ov2}.
}

On the other hand, subtracting the measurement-dependent part~\eqref{integral results} from the independent~\eqref{measurement-independent part}, we find that the semi-classical contribution to the Lee lower bound~\eqref{real for Lee inequality} vanishes:
\al{
\wt{\mc R}_{\st\rhoin}\fnl{\op x,\op p;\M_\tx{ph}}
	&=	0.
}

Unlike the LT inequality, the Lee inequality provides the meaningful lower bound, which has turned out to be identical to Heisenberg's, and is invariably saturated for the pure Gaussian initial state.

\subsection{Physical interpretation for uncertainty relations}\label{sec: Physical interpretation for uncertainty relations}
In the limit $\sigma\to0$ ($\infty$) both the LT and Lee errors for $\op x$ ($\op p$) go to zero.
As a result, for the LT errors, the product~\eqref{lhs for LT inequality} becomes zero, and the LT inequality~\eqref{LT inequality for Gaussian initial state} is saturated as $0=0$. In contrast, for the Lee errors, the product remains constant as in Eq.~\eqref{product of Lee errors}, even when one of the errors goes to zero, becuase the other diverges.

The vanishing product of LT errors in this limit can be interpreted in the context of the relation~\eqref{Ozawa and LT inequalities} to the Ozawa inequality as follows.
At an intermediate step (see Eq.~\eqref{Ozawa and LT inequalities in x,p}), the relation reads
\al{
		\varepsilon_{\st{\rhoin}}\fnl{\op{x}}
		\varepsilon_{\st{\rhoin}}\fnl{\op{p}}
	\sr{\tx{LT}}\geq
		\ub{\sqrt{
		\mc I_{\st\rhoin}^2\fnl{\op x,\op p}
		+\mc R_{\st\rhoin}^2\fnl{\op x,\op p}
		}}_{0}
	&\geq
		\ub{\ab{\mc I_{\st\rhoin}\fnl{\op x,\op p}}}_{0}\nn
	&\geq
		\ub{{1\ov2}\ab{\Braket{\commutator{\op x}{\op p}}_{\st{\rhoin}}}}_{{1\ov2}}
		-
			\ub{
				\ub{\varepsilon_{\st{\rhoin}}\fnl{\op x}}_{\sqrt{\sigma_\tx{red}\ov2}}
				\ub{\sigma_{\st\rhoin}\fnl{\op p}}_{{1\ov\sqrt{2\sigma_\tx{in}}}}
				}_{
				{1\ov2}\sqrt{\sigma\ov\sigma_\tx{sum}}
				}
		-
			\ub{
				\ub{\sigma_{\st\rhoin}\fnl{\op x}}_{\sqrt{\sigma_\tx{in}\ov2}}
				\ub{\varepsilon_{\st{\rhoin}}\fnl{\op p}}_{{1\ov\sqrt{2\sigma_\tx{sum}}}}
				}_{
				{1\ov2}\sqrt{\sigma_\tx{in}\ov\sigma_\tx{sum}}
				},
			\label{Ozawa and LT inequalities for Gaussian}
}
where we have abbreviated $\M_\tx{ph}$ for the errors, etc.

It is evident that the far right-hand side of Eq.~\eqref{Ozawa and LT inequalities for Gaussian} includes negative contributions that correspond to terms transposed from the left-hand side of the Ozawa inequality~\eqref{Ozawa inequality}. Importantly, the vanishing right-hand side of the LT inequality does not necessarily imply a vanishing lower bound for the Ozawa inequality.

Finally, we note that, as discussed in Sec.~\ref{Impossibility of vanishing LT errors for non-commuting observables}, even though the right-hand side of the LT inequality~\eqref{LT inequality for Gaussian initial state} vanishes, the inequality still forbids \emph{both} errors from vanishing simultaneously.

\section{Summary and discussion}\label{summary section}
We have demonstrated that the Gaussian wave-packet formalism can be a practical realization of the joint measurement of position and momentum, shedding light on Heisenberg's original philosophy of the uncertainty principle within the framework of Lee and Tsutsui's error, disturbance, and their uncertainty relations. We have successfully obtained the Lee-Tsutsui (LT) error and the refined Lee error in the context of position-momentum measurement for the first time. Our findings indicate that the LT uncertainty relation, in the limiting case of projective measurement of either position or momentum, becomes trivial: $0=0$.

In contrast, the refined Lee uncertainty relation, focusing on errors for local representability, provides a constant lower bound that remains unaffected by such limits and is invariably saturated for a pure Gaussian initial state. This lower bound aligns with Heisenberg's value. Our study delves into the Gaussian measurement process in the phase space, demonstrating its capability to interpolate smoothly between the projective measurements of position and momentum.

Building on the foundational work of Heisenberg; Kennard, Weyl, and Robertson; and Schr\"odinger, as well as on the advancements made by Arthurs, Kelly, and Goodman; Ozawa; and Watanabe, Sagawa, and Ueda, our research deepens the understanding of quantum uncertainty. We leverage the Gaussian wave-packet formalism to provide a comprehensive and tangible approach to the abstract constructs of the LT formalism. This paper advocates for the POVM measurement onto the Gaussian wave-packet basis as a natural method for joint position and momentum measurement, offering concrete expressions for previously abstract concepts like pullback and pushforward within the LT formalism. Our work bridges the theoretical and practical aspects of quantum mechanics, setting the stage for further research and applications in quantum uncertainty.

It has been shown that the Gaussian basis used in our current study on a fixed time slice can be extended to the whole spacetime~\cite{Ishikawa:2005zc,Ishikawa:2018koj,Ishikawa:2020hph,Ishikawa:2021bzf}; see Ref.~\cite{Oda:2021tiv} for the complete basis of Lorentz invariant wave packets and Ref.~\cite{Oda:2023qek} for that of Lorentz covariant ones including the spin degrees of freedom. It would be fascinating to explore the time-energy uncertainty relation on this ground.
It is also interesting to show the LT and Lee inequalities for the standard Clauser-Horne-Shimony-Holt (CHSH) inequality as a concrete realization in the finite-dimensional space.
These developments will be presented in subsequent publications.

\subsection*{Acknowledgment}
N.O.\ is indebted to Sho Machiyama for assistance in understanding the LT formalism and to Shogo Tanimura for fostering a welcoming research environment in his group. We are grateful to Gen Kimura for useful comments, Jaeha Lee for providing talk slides and offering useful comments, Kenji Nishiwaki for helpful conversations, and Juntaro Wada for useful comments. The work of K.O.\ is partially supported by JSPS KAKENHI Grant Nos.~JP19H01899 and JP21H01107.
We thank the Short-term Domestic Academic Exchange Fund of Tokyo Woman’s Christian University for financially supporting the stay of N.O.\ at the university, during which part of this paper was revised.

\appendix
\section*{Appendix}
In the Appendix, for readability, we write general operators as $\op x$ and $\op p$, both on $\mc H$, instead of $\op A$ and $\op B$. We do not assume that their commutator $\commutator{\op x}{\op p}$ takes any specific value, nor that the POVM $\opf\Pih{X,P}$ on $\mc H$ has the particular form~\eqref{Gaussian POVM}.
For this general POVM, we write the measurement as $\M$ instead of $\M_\tx{ph}$; see Eq.~\eqref{pdf in outcome space}.
Since only $\M$ appears in the Appendix, we omit $\M$ from errors, etc.

In the notation described above, the Ozawa inequality~\eqref{Ozawa inequality} takes the following forms:
\al{
\epsilon^\tx O_{\st\rh}\fnl{\op x}\epsilon^\tx O_{\st\rh}\fnl{\op p}
+\epsilon^\tx O_{\st\rh}\fnl{\op x}\sigma_{\st\rh}\fnl{\op p}
+\sigma_{\st\rh}\fnl{\op x}\epsilon^\tx O_{\st\rh}\fnl{\op p}
	&\geq
		{1\ov2}\ab{\Braket{\commutator{\op x}{\op p}}_{\st\rh}}.
		\label{Ozawa inequality in x,p}
}
We will demonstrate that the relation~\eqref{Ozawa and LT inequalities} between the LT and Ozawa inequalities follows from
\al{
		\varepsilon_{\st\rh}\fnl{\op x}
		\varepsilon_{\st\rh}\fnl{\op p}
	\sr{\tx{LT}}\geq
		\sqrt{
		\mc I_{\st\rh}^2\fnl{\op x,\op p}
		+\mc R_{\st\rh}^2\fnl{\op x,\op p}
		}
	&\geq
		\ab{\mc I_{\st\rh}\fnl{\op x,\op p}}\nn
	&\geq
		{1\ov2}\ab{\Braket{\commutator{\op x}{\op p}}_{\st\rh}}
		-\varepsilon_{\st\rh}\fnl{\op x}\sigma_{\st\rh}\fnl{\op p}
		-\sigma_{\st\rh}\fnl{\op x}\varepsilon_{\st\rh}\fnl{\op p},
			\label{Ozawa and LT inequalities in x,p}
}
based on the condition that
$\epsilon^\tx O_{\st\rh}\fnl{\op q}
	\geq	\varepsilon_{\st\rh}\fnl{\op q}$
for any observable $\op q$, as will be shown in Eq.~\eqref{Ozawa error is larger than LT error}.

\section{Ozawa inequality}\label{Ozawa section}
We review the derivation of Ozawa inequality~\eqref{Ozawa inequality in x,p}.

On the original space $\mc H$, we prepare a joint POVM $\opf\Pih{X,P}$ that satisfies
\al{
\int_{\R^2}\df X\,\df P\,\opf\Pih{X,P}
	&=	\op1,
}
as well as the positivity of the eigenvalues $0\leq\opf\Pih{X,P}\leq\op1$.
We also define the marginalized POVMs, on $\mc H$,
\al{
\opf\Pih[\op x]{X}
	&:=	\int_\R\df P\,\opf\Pih{X,P},&
\opf\Pih[\op p]{P}
	&:=	\int_\R\df X\,\opf\Pih{X,P}.
}

We prepare an ancilla space $\mc K$ and define on $\mc K$ a pair of “meter” observables, $\op X$ and $\op P$, which commute with each other:
\al{
\commutator{\op X}{\op P}=0.
	\label{X and P commute}
}
Without loss of generality, we may serve their simultaneous eigenvectors $\Ket{X,P}$ in $\mc K$:\footnote{
The simultaneous eigenvector $\Ket{X,P}$ in $\mc K$ and the Gaussian state $\Ket{X,P;\sigma}$ in $\mc H$ are distinct objects and must not be confused.
}
\al{
\op X\Ket{X,P}
	&=	X\Ket{X,P},&
\op P\Ket{X,P}
	&=	P\Ket{X,P},
}
with
\al{
\Braket{X,P|X',P'}
	&=	\delta\fn{X-X'}\delta\fn{P-P'},&
\int_{\R^2}\df X\,\df P\,\Ket{X,P}\Bra{X,P}
	&=	\op 1,
		\label{orthonormality of K basis}
}
that is,
\al{
\op X
	&=	\int_{\R^2}\df X\,\df P\,X\Ket{X,P}\Bra{X,P},&
\op P
	&=	\int_{\R^2}\df X\,\df P\,P\Ket{X,P}\Bra{X,P}.
	\label{X and P expanded}
}
We also define projection operators on $\mc K$ as
\al{
\opf\Eh[\op X]{X}
	&=	\int_\R\df P\,\Ket{X,P}\Bra{X,P},&
\opf\Eh[\op P]{P}
	&=	\int_\R\df X\,\Ket{X,P}\Bra{X,P},
}
where $\opf\Eh[\op X]{X}$ projects onto the eigenspace of $\op X$ with eigenvalue $X$.
In particular, we can show that
\al{
\commutator{\opf\Eh[\op X]{X}}{\opf\Eh[\op P]{P}}
	&=	0,&
\opf\Eh[\op X]{X}\opf\Eh[\op P]{P}
	&=	\Ket{X,P}\Bra{X,P}.
}

Ozawa claims that for any POVM $\opf\Pih{X,P}$, there exists a 5-tuple $\pn{\mc K,\st\xih,\op U,\op X,\op P}$ that realizes
\al{
\opf\Pih{X,P}
	&=	\Tr_{\mc K}\br{\op U[\dagger]\pn{\op1\ot\Ket{X,P}\Bra{X,P}}\op U\pn{\op1\ot\st\xih}}
	=:	\Ex{\op U[\dagger]\pn{\op1\ot\Ket{X,P}\Bra{X,P}}\op U}{\op1\ot\st\xih},
	\label{Ozawa's 5-tuple}
}
where $\Tr_{\mc K}$ is the partial trace over $\mc K$; $\op U$ is a unitary operator on $\mc H\ot\mc K$, typically representing a time translation; $\st\xih$ is a state (density operator) in $\mc K$; and $\mathbb E$ denotes what we refer to as the ``partial expectation value.'' In the Appendix, we adopt the convention that braces $\{\cdots\}$ indicate that the result is an operator in either subspace $\mc H$ or $\mc K$, unless they denote an anticommutator.

The outcome of the POVM measurement $\M$, corresponding to~\eqref{measurement onto phase space}, becomes\footnote{
In Ozawa's original paper~\cite{OZAWA2004367}, it is rather written in the form
\als{
\pdf{\sqbr{M\rh}}{X,P}
=	\Tr\fnl{
			\op U
			\pn{\st\rh\ot\st\xih}
			\op U[\dagger]
			\pn{\op1\ot\Ket{X,P}\Bra{X,P}}}
	=	\Braket{\op1\ot\Ket{X,P}\Bra{X,P}}_{\op U
			\pn{\st\rh\ot\st\xih}
			\op U[\dagger]}.
}
}
\al{
\pdf{\sqbr{M\rh}}{X,P}
	&=	\Tr_{\mc H}\Fnl{\opf\Pih{X,P}\st\rh}
	=	\Braket{\opf\Pih{X,P}}_{\st\rh}\nn
	&=	\Tr_{\mc H}\fnl{\Tr_{\mc K}\br{\op U[\dagger]\pn{\op1\ot\Ket{X,P}\Bra{X,P}}\op U\pn{\op1\ot\st\xih}}\st\rh}\nn
	&=	\Tr_{\mc H}\fnl{\Tr_{\mc K}\br{\op U[\dagger]\pn{\op1\ot\Ket{X,P}\Bra{X,P}}\op U\pn{\op1\ot\st\xih}\pn{\st\rh\ot\op1}}}\nn
	&=	\Tr\fnl{\op U[\dagger]\pn{\op1\ot\Ket{X,P}\Bra{X,P}}\op U\pn{\st\rh\ot\st\xih}}
	=	\Braket{
			\op U[\dagger]\pn{\op1\ot\Ket{X,P}\Bra{X,P}}\op U
			}_{\st\rh\ot\st\xih}\nn
	&=	\pdf{\sqbr{M\pn{\rh\ot\xih}}}{X,P},
			\label{pdf in outcome space}
}
where we used $\Tr_{\mc H}\fnl{\Tr_{\mc K}\!\br{\cdots}}=\Tr\fnl{\cdots}$, and the last expression is given by definition of the measurement in the LT formalism with the 5-tuple~\eqref{Ozawa's 5-tuple}.\footnote{
Abuse of notation should be understood without confusion: the map $\M$ in the first and last lines is from $\st\rh$ and from $\st\rh\ot\st\xih$, respectively.
}
Ozawa claims that after the measurement $\M$, the input state $\st\rh\in\stf Z{\mc H}$ is transformed into the conditional output state in $\mc H$,
\al{
\stf\rhoout{X,P}
	&=	{\Tr_{\mc K}\br{
			\op U
			\pn{\st\rh\ot\st\xih}
			\op U[\dagger]
			\pn{\op1\ot\Ket{X,P}\Bra{X,P}}}\ov
		\pdf{\sqbr{M\rh}}{X,P}
		},
}
although this point will not be used in the following discussion.

We define, on $\mc H\ot\mc K$,
\al{
\op\Xds
	&:=	\op U[\dagger]\pn{\op1\ot\op X}\op U,&
\op\Pds
	&:=	\op U[\dagger]\pn{\op1\ot\op P}\op U.
}
They commutes with each other by the assumption~\eqref{X and P commute}:
\al{
\commutator{\op\Xds}{\op\Pds}
	&=	0.
		\label{commutator of Xds and Pds}
}
By putting the expansion~\eqref{X and P expanded}, we can also write\footnote{\label{squared expressions due to diagonality}
Eq.~\eqref{Xds and Pds squared explicit} directly follows from our choice of the basis $\Ket{X,P}$ of $\mc K$ to be projective, as shown in Eq.~\eqref{orthonormality of K basis}, unlike the POVM $\opf\Pih{X,P}$ on $\mc H$.
}
\al{
\op\Xds
	&=	\int_{\R^2}\df X\,\df P\,X\,\op U[\dagger]\pn{\op1\ot\Ket{X,P}\Bra{X,P}}\op U,&
\op\Pds
	&=	\int_{\R^2}\df X\,\df P\,P\,\op U[\dagger]\pn{\op1\ot\Ket{X,P}\Bra{X,P}}\op U,
	\label{Xds and Pds explicit}\\
\op\Xds[2]
	&=	\int_{\R^2}\df X\,\df P\,X^2\,\op U[\dagger]\pn{\op1\ot\Ket{X,P}\Bra{X,P}}\op U,&
\op\Pds[2]
	&=	\int_{\R^2}\df X\,\df P\,P^2\,\op U[\dagger]\pn{\op1\ot\Ket{X,P}\Bra{X,P}}\op U,
		\label{Xds and Pds squared explicit}
}
and their simultaneous eigenvectors, in $\mc H\ot\mc K$, are
\al{
\op U[\dagger]\Pn{\Ket\psi\ot\Ket{X,P}}
}
with any state $\Ket\psi\in\mc H$.
We see that their projection operators, on $\mc H\ot\mc K$, become
\al{
\opf\Eh[\op\Xds]{X}
	&=	\int_\R\df P\,
		\sum_{\psi\tx{: complete set}}\op U[\dagger]\Pn{\Ket\psi\ot\Ket{X,P}}\Pn{\Bra\psi\ot\Bra{X,P}}\op U
	=	\op U[\dagger]\pn{\op1\ot\opf\Eh[\op X]{X}}\op U,\nn
\opf\Eh[\op\Pds]{P}
	&=	\int_\R\df X\,
		\sum_{\psi\tx{: complete set}}\op U[\dagger]\Pn{\Ket\psi\ot\Ket{X,P}}\Pn{\Bra\psi\ot\Bra{X,P}}\op U
	=	\op U[\dagger]\pn{\op1\ot\opf\Eh[\op P]{P}}\op U.
}
Again, we obtain
\al{
\commutator{\opf\Eh[\op\Xds]{X}}{\opf\Eh[\op\Pds]{P}}
	&=	0,&
\opf\Eh[\op\Xds]{X}\opf\Eh[\op\Pds]{P}
	&=	\op U[\dagger]\pn{\op1\ot\Ket{X,P}\Bra{X,P}}\op U.
}
Now we can rewrite the POVM~\eqref{Ozawa's 5-tuple} as
\al{
\opf\Pih{X,P}
	&=	\Tr_{\mc K}\br{\opf\Eh[\op\Xds]{X}\opf\Eh[\op\Pds]{P}\pn{\op1\ot\st\xih}}
	=	\Ex{\opf\Eh[\op\Xds]{X}\opf\Eh[\op\Pds]{P}}{\op1\ot\st\xih},
	\label{ancilla quadruple}
}
and the outcome pdf~\eqref{pdf in outcome space} as
\al{
\pdf{\sqbr{M\rh}}{X,P}
	&=	\Tr\fnl{\opf\Eh[\op\Xds]{X}\opf\Eh[\op\Pds]{P}\pn{\st\rh\ot\st\xih}}
	=	\Braket{\opf\Eh[\op\Xds]{X}\opf\Eh[\op\Pds]{P}}_{\st\rh\ot\st\xih}.
		\label{outcome in quadruple representation}
}
Ozawa calls any quadruple $\pn{\mc K,\st\xih,\op\Xds,\op\Pds}$ satisfying Eq.~\eqref{ancilla quadruple} the ancilla for the joint POVM $\opf\Pih{X,P}$.

The noise and mean noise operators are defined on $\mc H\ot\mc K$ and $\mc H$, respectively, as
\al{
\op N_{\op q}
	&:=	\op\Xds-\op q\ot\op 1,
	\label{Noise defined}\\
\op n_{\op q}
	&:=	\Tr_{\mc K}\br{\op N_{\op q}\pn{\op1\ot\st\xih}}
	=	\Ex{\op N_{\op q}}{\op1\ot\st\xih},
		\label{mean noise operators defined}
}
for $\op q=\op x,\op p$.
The first and second moment operators are defined on $\mc H$ as
\al{
\opf\Oh{\op\Pi[\op x]}
	&:=	\int_\R\df X\,X\,\opff\Pih[\op x]X,&
\opf\Oh{\op\Pi[\op p]}
	&:=	\int_\R\df P\,P\,\opff\Pih[\op p]P,\\
\opf\Oh[(2)]{\op\Pi[\op x]}
	&:=	\int_{\R^2}\df X\,X^2\opff\Pih[\op x]X,&
\opf\Oh[(2)]{\op\Pi[\op p]}
	&:=	\int_{\R^2}\df P\,P^2\opff\Pih[\op p]P.
}
These can be rewritten as
\al{
\opf\Oh{\op\Pi[\op x]}
	&=	\br{\op U[\dagger]\pn{\op1\ot\op X}\op U}_{\op1\ot\st\xih},&
\opf\Oh{\op\Pi[\op p]}
	&=	\br{\op U[\dagger]\pn{\op1\ot\op P}\op U}_{\op1\ot\st\xih},\\
\opf\Oh[(2)]{\op\Pi[\op x]}
	&=	\br{\op U[\dagger]\pn{\op1\ot\op X[2]}\op U}_{\op1\ot\st\xih},&
\opf\Oh[(2)]{\op\Pi[\op p]}
	&=	\br{\op U[\dagger]\pn{\op1\ot\op P[2]}\op U}_{\op1\ot\st\xih}.
				\label{second moments rewritten}
}
It follows that on $\mc H$,
\al{
\op n_{\op q}
	&=	\opf\Oh{\op\Pi[\op q]}-\op q
	\label{Ozawa's relation for the first moment}
}
for $\op q=\op x,\op p$.

Now the Ozawa error is defined by
\al{
\epsilon^\tx O_{\st\rh}\fnl{\op q}
	&:=	\abb{\op N_{\op q}}_{\st\rh\ot\st\xih}
	=	\sqrt{\Tr\fnl{\opnh\Nh[2][\op q]\pn{\st\rh\ot\st\xih}}}
		\label{Ozawa error defined}
}
for $\op q=\op x,\op p$.
It follows that
\al{
\epsilon^\tx O_{\st\rh}\fnl{\op q}
	&=	\sqrt{\Tr_{\mc H}\fnl{\pn{
			\opf\Oh[(2)]{\op\Pi[\op q]}
			-\opf\Oh[2]{\op\Pi[\op q]}
			+\opnh\nh[2][\op q]
			}\st\rh}},\\
\sigma_{\st\rh\ot\st\sh}^2\fnl{\op N_{\op q}}
	&=	\pn{\epsilon^\tx O_{\st\rh}\fnl{\op q}}^2
		-\pn{\Tr_{\mc H}\fnl{\opnh\nh_{\op q}\st\rh}}^2.
}

Putting Eq.~\eqref{Noise defined} into the commutator~\eqref{commutator of Xds and Pds},
we obtain
\al{
0	&=
		\commutator{\op N_{\op x}}{\op N_{\op p}}
		+\commutator{\op N_{\op x}}{\op p\ot\op1}
		+\commutator{\op x\ot\op1}{\op N_{\op p}}
		+\commutator{\op x}{\op p}\ot\op1.
}
Therefore,
\al{
\Braket{\commutator{\op N_{\op x}}{\op N_{\op p}}}_{\st\rh\ot\st\xih}
+\Braket{\commutator{\op N_{\op x}}{\op p\ot\op1}}_{\st\rh\ot\st\xih}
+\Braket{\commutator{\op x\ot\op1}{\op N_{\op p}}}_{\st\rh\ot\st\xih}
	&=-\Braket{\commutator{\op x}{\op p}\ot\op1}_{\st\rh\ot\st\xih}\nn
	&=-\Braket{\commutator{\op x}{\op p}}_{\st\rh},
}
and hence
\al{
\ab{\Braket{\commutator{\op N_{\op x}}{\op N_{\op p}}}_{\st\rh\ot\st\xih}}
+\ab{\Braket{\commutator{\op N_{\op x}}{\op p\ot\op1}}_{\st\rh\ot\st\xih}}
+\ab{\Braket{\commutator{\op x\ot\op1}{\op N_{\op p}}}_{\st\rh\ot\st\xih}}
	&\geq
		\ab{\Braket{\commutator{\op x}{\op p}}_{\st\rh}}.
}
Applying the KR inequality
\al{
\sigma_{\st\rh}\fnl{\op A}\sigma_{\st\rh}\fnl{\op B}\geq{1\ov2}\ab{\Braket{\commutator{\op A}{\op B}}_{\st\rh}}
	\label{KR inequality}
}
for each term, we get
\al{
\sigma_{\st\rh\ot\st\xih}\fnl{\op N_{\op x}}
	\sigma_{\st\rh\ot\st\xih}\fnl{\op N_{\op p}}
+\sigma_{\st\rh\ot\st\xih}\fnl{\op N_{\op x}}
	\ub{\sigma_{\st\rh\ot\st\xih}\fnl{\op p\ot\op1}}_{\sigma_{\st\rh}\fnl{\op p}}
+\ub{\sigma_{\st\rh\ot\st\xih}\fnl{\op x\ot\op1}}_{\sigma_{\st\rh}\fnl{\op x}}
	\sigma_{\st\rh\ot\st\xih}\fnl{\op N_{\op p}}
	&\geq
		{1\ov2}\ab{\Braket{\commutator{\op x}{\op p}}_{\st\rh}},
}
where we have used the following operator identity on $\mc H$:
$\Tr_{\mc K}\Br{\op A\Pn{\op c\ot\op 1}}
	=
		\Tr_{\mc K}\Br{\op A}\op c$,
in which $\op A$ and $\op c$ are arbitrary operators on $\mc H\ot\mc K$ and on $\mc H$, respectively.
Since we have
\al{
\abb{\op A}_{\st\rh}^2
	=	\sigma_{\st\rh}^2\fnl{\op A}
		+\Braket{\op A}_{\st\rh}^2
	\geq
		\sigma_{\st\rh}^2\fnl{\op A},
}
it follows that
\al{
\abb{\op N_{\op x}}_{\st\rh\ot\st\xih}
	\abb{\op N_{\op p}}_{\st\rh\ot\st\xih}
+\abb{\op N_{\op x}}_{\st\rh\ot\st\xih}
	\sigma_{\st\rh}\fnl{\op p}
+\sigma_{\st\rh}\fnl{\op x}
	\abb{\op N_{\op p}}_{\st\rh\ot\st\xih}
	&\geq
		{1\ov2}\ab{\Braket{\commutator{\op x}{\op p}}_{\st\rh}}.
}
By applying the definition~\eqref{Ozawa error defined}, we establish the Ozawa inequality~\eqref{Ozawa inequality in x,p}.

\section{Relation between Ozawa and LT inequalities}\label{Ozawa and LT section}
We review how the LT inequality~\eqref{LT inequality} serves as a necessary condition for the Ozawa inequality~\eqref{Ozawa inequality in x,p}, as shown in Eq.~\eqref{Ozawa and LT inequalities in x,p}.\footnote{
The relationship between the Ozawa inequality~\cite{OZAWA2004367} and the LT inequality~\cite{lee2023universalformulationuncertaintyrelation}, both for noise and disturbance, can also be treated on equal footing.
}

From the outcome pdf given in Eq.~\eqref{pdf in outcome space} or \eqref{outcome in quadruple representation}, we obtain, on $\mc H$, the LT adjoint of a function~$\rf f$ as
\al{
\op{\la Mf}
	&=	\int_{\R^2}\df X\,\df P\,\rff f{X,P}\,\opf\Pih{X,P},
		\label{pullback on H}
}
and on $\mc H\ot\mc K$ as
\al{
\op{\la Mf}
	&=	\int_{\R^2}\df X\,\df P\,\rff f{X,P}\,
		\op U[\dagger]\pn{\op1\ot\Ket{X,P}\Bra{X,P}}\op U.
}
Although we use the same notation for the LT adjoint on $\mc H$ and on $\mc H\ot\mc K$, the distinction should be clear from the context.

Putting the LT adjoint~\eqref{pullback on H} into Eq.~\eqref{definition of pushforward} with Ozawa's 5-tuple~\eqref{Ozawa's 5-tuple}, we find the pushforward function of an observable $\op q$ on $\mc H$, in a parallel manner to Eq.~\eqref{pdf in outcome space}:
\al{
\rff{\sqbr{M_{\rh\star}\wh q}}{X,P}
	&=	\Braket{\op q,\,\opf\Pih{X,P}}_{\st\rh}
	=	\Tr_{\mc H}\fnl{{\anticommutator{\op q}{\opf\Pih{X,P}}\ov2}\st\rh}\nn
	&=	\Tr\fnl{{\anticommutator{\op q\ot\op1}{\op U[\dagger]\pn{\op1\ot\Ket{X,P}\Bra{X,P}}\op U}\ov2}\pn{\st\rh\ot\st\xih}}\nn
	&=	\Braket{\op q\ot\op1,\,\op U[\dagger]\pn{\op1\ot\Ket{X,P}\Bra{X,P}}\op U}_{\st\rh\ot\st\xih}
	=	\rff{\sqbr{M_{\pn{\rh\ot\xih}\star}\pn{\wh q\ot\wh1}}}{X,P}.
	\label{relation between pushforward functions}
}
We see that the pushforward function from $\op q$ coincides with that from $\op q\ot\op1$.

From Eq.~\eqref{LT adjoint defined}, we can read off
\al{
\rff{\sqbr{M^{\star-1}\wh\Xds}}{X,P}
	&=	X,&
\rff{\sqbr{M^{\star-1}\wh\Pds}}{X,P}
	&=	P,
		\label{inverse of LT adjoint for Xds and Pds}
}
as follows: By substituting Eq.~\eqref{Xds and Pds explicit}, we obtain
\al{
\Braket{\op\Xds}_{\st\rh\ot\st\xih}
	&=	\int_{\R^2}\df X\,\df P\,X
			\ub{
				\Tr\fnl{\op U[\dagger]\pn{\op1\ot\Ket{X,P}\Bra{X,P}}\op U\pn{\st\rh\ot\st\xih}}
				}_{
				\pdf{M\pn{\rh\ot\xih}}{X,P}
				},
}
which is to be compared with
\al{
\Braket{\rf{M^{\star-1}\wh\Xds}}_{\pd{M\pn{\rh\ot\xih}}}
	&=	\int_{\R^2}\df X\,\df P\,
			\rff{\sqbr{M^{\star-1}\wh\Xds}}{X,P}
			\pdf{\sqbr{M\pn{\rh\ot\xih}}}{X,P};
}
and similarly for $P$. Thus, we arrive at Eq.~\eqref{inverse of LT adjoint for Xds and Pds}.

It follows that
\al{
\abb{\op\Xds}_{\st\rh\ot\st\xih}^2
	&=	\int_{\R^2}\df X\,\df P\,X^2\Tr\fnl{\op U[\dagger]\Pn{\op1\ot\Ket{X,P}\Bra{X,P}}\op U\pn{\st\rh\ot\st\xih}}_{\st\rh\ot\st\xih}
	=	\abb{\rf{M^{\star-1}\wh\Xds}}_{\pd{M\pn{\rh\ot\xih}}}^2,\nn
\abb{\op\Pds}_{\st\rh\ot\st\xih}^2
	&=	\int_{\R^2}\df X\,\df P\,P^2\Tr\fnl{\op U[\dagger]\Pn{\op1\ot\Ket{X,P}\Bra{X,P}}\op U\pn{\st\rh\ot\st\xih}}_{\st\rh\ot\st\xih}
	=	\abb{\rf{M^{\star-1}\wh\Pds}}_{\pd{M\pn{\rh\ot\xih}}}^2,
		\label{equality of norms}
}
where we used Eq.~\eqref{Xds and Pds squared explicit} in the first step and Eqs.~\eqref{pdf in outcome space} and \eqref{inverse of LT adjoint for Xds and Pds} in the last step.

Having prepared the above, we now come to the derivation of the relation~\eqref{Ozawa and LT inequalities in x,p}. First, its leftmost inequality can be shown as follows:
\al{
\pn{\epsilon^\tx O_{\st\rh}\fnl{\op x}}^2-\varepsilon_{\st\rh}^2\fnl{\op x}
	&=	\abb{\op\Xds-\op x\ot\op1}_{\st\rh\ot\st\xih}^2
		-\pn{\abb{\op x}_{\st\rh}^2
			-\abb{\rf{M_{\rh\star}\wh x}}_{\pd{M\rh}}^2}\nn
	&=	\abb{\op\Xds}_{\st\rh\ot\st\xih}^2
		-2\ub{
			\Braket{\op\Xds,\op x\ot\op1}_{\st\rh\ot\st\xih}
			}_{
			\Braket{\rf{M^{\star-1}\wh\Xds},\,\rf{M_{\pn{\rh\ot\xih}\star}\pn{\wh x\ot\wh1}}}_{\pd{M\pn{\rh\ot\xih}}}
			}
		+\ub{\abb{\op x\ot\op1}^2_{\st\rh\ot\st\xih}
			-\abb{\op x}^2_{\st\rh}}_{
			0}
		+\ub{
			\abb{\rf{M_{\rh\star}\wh x}}_{\pd{M\rh}}^2
			}_{
			\abb{\rf{M_{\pn{\rh\ot\xih}\star}\pn{\wh x\ot\wh1}}}^2_{\pd{M\pn{\rh\ot\xih}}}
			}\nn
	&=	\ub{\abb{\op\Xds}_{\st\rh\ot\st\xih}^2
			-\abb{\rf{M^{\star-1}\wh\Xds}}_{\pd{M\pn{\rh\ot\xih}}}^2}
			_{
			0}
		+\abb{\rf{M^{\star-1}\wh\Xds}-\rf{M_{\pn{\rh\ot\xih}\star}\pn{\wh x\ot\wh1}}}_{\pd{M\pn{\rh\ot\xih}}}^2
	\geq
		0,
		\label{Ozawa error is larger than LT error}
}
where we have used Eq.~\eqref{equality of norms} in the last step; and similarly for $\op p$.

Next, the rightmost inequality in Eq.~\eqref{Ozawa and LT inequalities in x,p} can be shown as follows:\footnote{
We thank Jaeha Lee for the useful comment on this part.
}
\al{
\ab{\mc I_{\st\rh}\fnl{\op x,\op p}}
	&=	\ab{\Braket{
			\commutator{\op x-\op{\la MM_\star \wh x}}{\op p}\ov2i
			}_{\st\rh}
		+\Braket{\commutator{\op x}{\op p-\op{\la MM_\star \wh p}}\ov2i}_{\st\rh}
		-\Braket{\commutator{\op x}{\op p}\ov2i}_{\st\rh}}\nn
	&\geq
		\ab{\Braket{\commutator{\op x}{\op p}\ov2i}_{\st\rh}}
		-\ab{\Braket{
			\commutator{\op x-\op{\la MM_\star \wh x}}{\op p}\ov2i
			}_{\st\rh}}
		-\ab{\Braket{\commutator{\op x}{\op p-\op{\la MM_\star \wh p}}\ov2i}_{\st\rh}
			}\nn
	&\geq
		\ab{\Braket{\commutator{\op x}{\op p}\ov2i}_{\st\rh}}
		-\sigma_{\st\rh}\fnl{\op x-\op{\la MM_\star \wh x}}
			\sigma_{\st\rh}\fnl{\op p}
		-\sigma_{\st\rh}\fnl{\op x}
			\sigma_{\st\rh}\fnl{\op p-\op{\la MM_\star \wh p}},
\label{rightmost inequality}
}
where the KR inequality~\eqref{KR inequality} was used in the last step. Then:
\al{
\sigma_{\st\rh}^2\fnl{\op A-\op{M^\star M_\star\op A}}
	&=
		\abb{\op A-\op{M^\star M_\star\op A}}_{\st\rh}^2
		-\Braket{\op A-\op{M^\star M_\star\op A}}_{\st\rh}^2\nn
	&=	\abb{\op A}_{\st\rh}^2
		-2\ \ub{\Braket{\op A,\,\op{M^\star M_\star\op A}}_{\st\rh}}_{\abb{\rf{M_\star\wh A}}_{\pd{M\rh}}^2}
		+\abb{\op{M^\star M_\star\op A}}_{\st\rh}^2
		-\Braket{\op A-\op{M^\star M_\star\op A}}_{\st\rh}^2\nn
	&=	\varepsilon_{\st\rh}^2\fnl{\op A}
		\ub{
			-\abb{\rf{M_\star\wh A}}_{\pd{M\rh}}^2
			+\abb{\op{M^\star M_\star\op A}}_{\st\rh}^2
			}_{\leq0}
		-\Braket{\op A-\op{M^\star M_\star\op A}}_{\st\rh}^2
	\leq
		\varepsilon_{\st\rh}^2\fnl{\op A},
		\label{a variance vs Ozawa error}
}
where the definition~\eqref{LT error defined} and inequality~\eqref{inequality for LT adjoint} were used in the second-last and final steps, respectively.
Substituting Eq.~\eqref{a variance vs Ozawa error} into Eq.~\eqref{rightmost inequality}, and then applying the inequality~\eqref{Ozawa error is larger than LT error}, we obtain the rightmost inequality in Eq.~\eqref{Ozawa and LT inequalities in x,p}.

\bibliographystyle{JHEP}
\bibliography{refs}

\end{document}